\documentclass[notitlepage]{revtex4-1}
\usepackage{amssymb}
\usepackage{amsfonts}
\usepackage{amsmath}
\usepackage{graphicx}
\usepackage{enumerate}
\usepackage{ulem}%
\usepackage{color}
\newcommand{\red}[1]{\textcolor{red}{#1}}

\providecommand{\U}[1]{\protect\rule{.1in}{.1in}}
\begin{document}
\title{Classical Density functional theory, unconstrained crystallization and polymorphic behaviour}
\author{James F. Lutsko}
\author{Julien Lam}
\affiliation{Center for Nonlinear Phenomena and Complex Systems, Code Postal 231,
Universit\'{e} Libre de Bruxelles, Blvd. du Triomphe, 1050 Brussels, Belgium}
\email{jlutsko@ulb.ac.be}
\homepage{http://www.lutsko.com}

\pacs{64.60.Q-, 82.60.Nh, 05.40.-a}
\begin{abstract}
  While in principle, classical density functional theory (cDFT) should be a powerful tool for the study of crystallization, in practice this has not so far been the case. Progress has been hampered by technical problems which have plagued the study of the crystalline systems using the most sophisticated Fundamental Measure Theory  models. In this paper, the reasons for the difficulties are examined and it is proposed that the tensor functionals currently favored are in fact numerically unstable. By reverting to an older, more heuristic model it is shown that all of the technical difficulties are eliminated. Application to a Lennard-Jones fluid results in a demonstration of power of cDFT to describe crystallization in a highly inhomogeneous system. First, we show that droplets attached to a slightly hydrophobic wall crystallize spontaneously upon being quenched. The resulting crystallites are clearly faceted structures and are predominantly HCP structures. In contrast, droplets in a fully periodic calculational cell remain stable to lower temperatures and eventually show the same spontaneous localization of the density into ``atoms'' but in an amorphous structure having many of the structural charactersitics of a glass. A small change of the protocol leads, at the same temperature, to the formation of crystals, this time with the FCC structure typical of bulk Lennard-Jones solids. The FCC crystals have lower free energy than the amorphous structures which in turn are more stable than the liquid droplets. It is demonstrated that as the temperature is raised, the free energy differences between the structures decreases until the solid clusters become less stable than the liquid droplets and spontaneously melt. The presence of energy barriers separating the various structures is therefore clearly demonstrated.
\end{abstract}
\date{\today }
\maketitle

\section{Introduction}

{Work on classical Density Functional Theory (cDFT) for classical systems\cite{Evans79,LutskoAdvChemPhysDFT} has long been driven by the promise of having a tool that allows for the determination of free energies and inhomogeneous density distributions for arbitary systems such as confined liquids and  solids and glasses (confined or not) and the energy barriers separating these states that would then open the way to studying all aspects of nucleation: lifetimes of metastable states, transition pathways etc without problem-specific modelling bias. Indeed, these goals are so important that they have driven much recent work in other fields as well. A good example is the field of computer simulation where the search for unbiased methods of studying free energy surfaces and transitions between metastable states that do not rely on the definition of collective variables to characterize such states has been prominant (see, e.g., the contributions of Piaggi et al\cite{Piaggi} and of Swinburn and Marinica\cite{Swinburne} that present two recent steps in this direction). In principle, cDFT evades the need for collective variables since one works directly with the density field. A well-known and closely related alternative to cDFT are Phase-Field theories which are a more mesoscopic theoretical approach that extends the Landau theory of phase transitions by adding additional variables and terms to the gradient-expanded free energy as suggested from DFT (for an overview, see e.g. \cite{PhaseField} where it is emphasized that phase-field theory can perhaps best be viewed as a mesoscopic, coarse-grained version of cDFT). But the truely microscopic approach based on cDFT that is capable of accurately describing atomic-scale correlations in liquids as well as solids and amorphous systems has so far remained elusive.

All forms density functional theory (classical and quantum, zero and finite temperature) are formally based on
exact theorems stating that for a fixed temperature and chemical potential in
the grand canonical ensemble, there exists a functional $\Lambda\left[
n\right]  $ of the local number density $n\left(  \mathbf{r}\right)  $ (or
partial densities for a multicomponent system) having the property that its
value is uniquely minimized by the equilibrium density, i.e. if the
equilibrium density is $n_{eq}\left(  \mathbf{r}\right)  $ then $\Lambda
\left[  n\right]  \geq\Lambda\left[  n_{eq}\right]  $ with equality if and
only if $n\left(  \mathbf{r}\right)  =n_{eq}\left(  \mathbf{r}\right)  $ .
Furthermore, when evaluated at this minimizing density field, one has that
$\Lambda\left[  n_{eq}\right]  =\Omega$, the grand canonical free energy for
the system. The functional $\Lambda$ depends also on any external fields
present, but this dependence is trivial, being given by $\Lambda\left[
n;\phi\right]  =F\left[  n\right]  +\int\left(  \phi\left(  \mathbf{r}\right)
-\mu\right)  n\left(  \mathbf{r}\right)  d\mathbf{r}$ where $\phi\left(
\mathbf{r}\right)  $ is the external potential, $\mu$ is the chemical
potential and $F\left[  n\right]$, called the Helmholtz free energy functional,  is the part of $\Lambda$ independent of
the field. In general, $F\left[  n\right]  $ is not known and so applications
of cDFT depend on some sort of approximation for it. These are often
constructed based on the rare systems for which exact results are possible
such as the ideal gas, hard particles in highly confined geometries and
hard-rods in one dimension. For this reason, the most highly refined theories
in three dimensions are for hard-spheres and so it has long been the case
that a crucial test of model cDFT's has been their ability to describe
inhomogeneous hard-sphere systems, including the solid phase and the
liquid-solid transition, see e.g. \cite{Evans79,LutskoAdvChemPhysDFT}.

The most accurate and widely accepted models today are of a class collectively
known as Fundamental Measure Theory (FMT)\cite{rosenfeld1}. These arose out of the exact
results of Percus and coauthors for one-dimensional systems\cite{Percus1,Percus2,Percus3} as well as an
approximate description of hard-sphere statistical mechanics called Scaled
Particle Theory\cite{spt1}. Rosenfeld's original FMT gave a good description of fluids
near a wall but failed to stabilize (i.e. to have a minimum corresponding to)
the hard-sphere solid phase. After much work by several authors, Tarazona
suggested a modification of Rosenfeld's theory that succeeded in stabilizing
the hard-sphere solid and that was motivated by demanding that the functional reproduce certain exact results\cite{tarazona-freezing}. A heuristic
modification of Tarazona's theory was proposed by Roth et al\cite{white_bear} that aimed to
improve the description of the dense liquid and this version of FMT, known as
the White Bear (WBI)\ functional, was for some time considered to be the best
available model. For example, it gives a very good quantitative prediction of
the hard-sphere freezing transition. A relatively recent modification, the
so-called White Bear II (WBII) model, gives an even more accurate
representation of the hard-sphere solid including a good description of the
vacancy concentration\cite{WBII}. These models are not perfect - there are difficulties
in extending them to mixtures\cite{Cuesta} and to metastable lattice structures\cite{LutskoBCC} - but their
value as the best current description of the hard-sphere solid cannot be questioned.

Originally, because of computational constraints, cDFT calculations for the
solid phase were performed using a resticted model of the density field
whereby it was represented as a sum of Gaussians centered at the fixed lattice
sites of a given Bravais lattice. Only the widths of the Gaussians and
sometimes their amplitudes, were varied so as to minimize $\Lambda$. This
clearly accords with an intuitive idea of what a solid should look like and
also has recovers the homogeneous liquid in the limit that the Gaussian widths
become infinetly large. Non-Gaussian corrections were also sometimes
investigated but generally seemed of minor importance. Only in the last 20
years have full-scale, uncontrained cDFT calculations been performed using
either finite element or pseudo-spectral methods. In particular, Oettel et al
have pioneered the investigation of the hard-sphere solid phase with finite
element methods and using the WB models\cite{Oettel}. However, one surprising point
noted in the course of these applications has been a numerical delicacy of the
WB models. In order to obtain useful results and to avoid numerical divergences, the authors report having to  work at constant number rather than constant chemical potential, as is more natural in the cDFT framework and in some cases being forced to carefully maintain strict cubic symmetry in the course of the numerical
optimizations as well as having to use relatively fine numerical discretizations.

Because of these limitations, the results reported so far on the use of cDFT to describe the solid phase has generally been limited to homogeneous solids, to planar liquid-solid interfaces and to two-dimensional systems.  For example, the homogeneous hard-sphere crystal has long been  used as a test of cDFT models and, less often, the phase diagram of homogeneous solids for systems with pair potentials have been studied(see, e.g., the discussion in \cite{lutsko:acp}). But in all cases, these calculations are based on pre-defined lattice structures and the creation ``by hand'' of localized density fields. One of the few applications beyond homogeneous crystals that have been studied in the literature by many authors is the planar liquid-solid hard-sphere interface, e.g. by Curtin\cite{Curtin}, Lutsko\cite{LutskoWall} and Oettel and coworkers\cite{OettelWall1,OettelWall2},  but always starting with the introduction of pre-selected crystalline structures by hand. Similarly, cDFT has long been applied to study the free energy of glass-like systems \cite{Baus, Lubchenko} but, once again, based on pre-determined structures (typically the pair-distribution function of Bennett\cite{Bennett1}). In contrast, in our work, we demonstrate the spontaneous formation of solid clusters having no relation at all to the symmetries of the calculational cell or of the applied boundary conditions and we {\em derive} a pair-distribution function from our stuctures that is the equivalent of the Bennett result. Only a few earlier works have demonstrated any spontaneous localization of a density field (e.g. the work on hard-sphere glasses of Dasgupta and Walls\cite{Dasgupta} and the extension of this work by Chaudhuri et al\cite{PhysRevLett.100.125701} which used very simple models limited to homogeneous systems,  and that of Archer and coworkers\cite{Archer1, Archer2} on two-dimensional freezing of soft-matter) and we are not aware of any such work applied to highly inhomogeneous systems such as clusters (as is necessary for applications like clustering in proteins, nucleation and crystallization) in three dimensions.

In this paper, we propose that the numerical difficulties of the WB functionals are
due to a pathology of the tensor models: namely, that their approximation for
$\Lambda\left[  n\right]  $  is not \red{bounded from below} and that in fact it has
no minimum. In this interpretation, the constraints mentioned above - cubic
symmetry and constant number - suppress this pathology to allow results to be
obtained from (artificial or actual) metastable stationary points of the functional. But while the constant number constraint is benign, the constraint
of cubic symmetry clearly limits applicability of the model to high symmetry
systems and thus restricts the overall utility of the model. In this paper, it
is shown that using a more heuristic, but demonstrably \red{stable} 
model, eliminates all of these problems and leads to a much more robust
description of the solid. The resulting model is not a definitive replacement
for the tensor models - the exact limits satisfied by the latter are
undoubtedly crucial to incorporate - but serves as a proof of concept of the
feasibility of a fully robust cDFT solid.

We go on to use this robust hard-sphere functional to construct a functional for arbitrary pair potentials by treating the attractive tail of the potential using the standard mean-field approach. This model is used to perform calculations for a Lennard-Jones potential. We first determine the entire vapor-liquid-solid phase diagram. Next, we study the behaviour of small droplets in contact with a wall. As the temperature is quenched, typical layering structure develops within the droplets until at a certain point, the density field {\em spontaneously} localizes into atoms arranged in a faceted, HCP crystallite. The same protocol applied to the case of spherical droplets with no walls present shows a similar spontaneous localization but into an {\em amorphous} structure with many of the structural characteristics found in a glass. An intermediate protocol where a ``wall'' is present at the start of minimization but then removed results in the formation of homogeneous crystals which are again faceted but which have the FCC structure typical of bulk Lennard-Jones. We show that at intermediate temperatures, all three structures - droplet, crystal and amorphous cluster -  are metastable minima of the free energy functional thus implying the presence of energy barriers separating them. These results clearly show the utility and far-reaching protential of this model for the study of nucleation, crystallization, polymorphism and, potentially, the glass transition.

Section II of this paper describes the technical problems with current hard-sphere functionals and our proposal for their origin and resolution. The result is a functional that is more numerically robust than any of the previous models. We implemented this functional in a new  finite-element algorithm (described in detail in an appendix) designed to rigorouly maintain the term-by-term positivity of the contributions to the $\Lambda$-functional and validate it by detailed comparison to previous works on hard-spheres. In Section III, the new hard-sphere functional is used to construct a mean-field description of a system interacting via arbitrary pair-potentials and applied specifically to a Lennard-Jones potential. Calculations are presented for the phase diagram of the homogeneous system and of critical clusters for the vapor-liquid and vapor-solid transitions.  We find that the solid clusters exhibit polymorphism depending on the environment in which they form. Section IV summarizes the results and discusses their implications for futher improvements of FMT hard-sphere models and for applications to the study of clustering, nucleation, crystallization and amorphous systems.

\section{Fundamental Measure Theory}
\subsection{FMT Models}
In this and the following Sections, attention will be restricted to single component systems in
three dimensions composed of hard spheres with diameter $\sigma$. The
Helmholtz free energy functional is written as $F\left[  n\right]  =$
$F_{id}\left[  n\right]  +F_{ex}\left[  n\right]  $ with the ideal gas
contribution given by
\begin{equation}
F_{id}\left[  n\right]  =k_{B}T\int\left\{  n\left(  \mathbf{r}\right)
\ln\left(  n\left(  \mathbf{r}\right)  \sigma^{3}\right)  -n\left(
\mathbf{r}\right)  \right\}  d\mathbf{r}%
\end{equation}
where $k_{B}$ is Boltzmann's constant and $T$ is the temperature and, for later use, we define $\beta = 1/(k_{B}T)$. The second
term is the excess free energy and in Fundamental Measure Theory is modeled
using Rosenfeld's ansatz as
\begin{equation}
F\left[  n\right]  =\int\Phi\left(  n_{\alpha}\left(  \mathbf{r;}\left[
n\right]  \right)  \right)  d\mathbf{r}%
\end{equation}
where $n_{\alpha}\left(  \mathbf{r;}\left[  n\right]  \right)  $ stands for a
family of linear functionals of the density having the generic form%
\begin{equation}
n_{\alpha}\left(  \mathbf{r;}\left[  n\right]  \right)  =\int w_{\alpha
}\left(  \left\vert \mathbf{r}-\mathbf{r}^{\prime}\right\vert \right)
n\left(  \mathbf{r}^{\prime}\right)  d\mathbf{r}^{\prime}%
\end{equation}
and the weights $w_{\alpha}\left(  r\right)  $ are in general short-ranged
functions. The standard density functionals are the local packing fraction,
\begin{equation}
\eta\left(  \mathbf{r}\right)  =\int\Theta\left(  \frac{\sigma}{2}-\left\vert
\mathbf{r}-\mathbf{r}^{\prime}\right\vert \right)  n\left(  \mathbf{r}%
^{\prime}\right)  d\mathbf{r}^{\prime},
\end{equation}
which is just the density averaged over a spherical volume corresponding to the
hard-sphere diameter, and the scalar, vector and tensor surface averages%
\begin{align}
s\left(  \mathbf{r}\right)   &  =\int\delta\left(  \frac{\sigma}{2}-\left\vert
\mathbf{r}-\mathbf{r}^{\prime}\right\vert \right)  n\left(  \mathbf{r}%
^{\prime}\right)  d\mathbf{r}^{\prime}\\
\mathbf{v}\left(  \mathbf{r}\right)   &  =\int\frac{\mathbf{r}-\mathbf{r}%
^{\prime}}{\left\vert \mathbf{r}-\mathbf{r}^{\prime}\right\vert }\delta\left(
\frac{\sigma}{2}-\left\vert \mathbf{r}-\mathbf{r}^{\prime}\right\vert \right)
n\left(  \mathbf{r}^{\prime}\right)  d\mathbf{r}^{\prime}\nonumber\\
\mathbf{T}\left(  \mathbf{r}\right)   &  =\int\frac{\mathbf{r}-\mathbf{r}%
^{\prime}}{\left\vert \mathbf{r}-\mathbf{r}^{\prime}\right\vert }%
\frac{\mathbf{r}-\mathbf{r}^{\prime}}{\left\vert \mathbf{r}-\mathbf{r}%
^{\prime}\right\vert }\delta\left(  \frac{\sigma}{2}-\left\vert \mathbf{r}%
-\mathbf{r}^{\prime}\right\vert \right)  n\left(  \mathbf{r}^{\prime}\right)
d\mathbf{r}^{\prime}\nonumber
\end{align}
or, equivalently,%
\begin{align}
s\left(  \mathbf{r}\right)   &  =\int\delta\left(  \frac{\sigma}{2}-r^{\prime
}\right)  n\left(  \mathbf{r}-\mathbf{r}^{\prime}\right)  d\mathbf{r}^{\prime
}\\
\mathbf{v}\left(  \mathbf{r}\right)   &  =\int\frac{\mathbf{r}^{\prime}%
}{r^{\prime}}\delta\left(  \frac{\sigma}{2}-r^{\prime}\right)  n\left(
\mathbf{r}-\mathbf{r}^{\prime}\right)  d\mathbf{r}^{\prime}\nonumber\\
\mathbf{T}\left(  \mathbf{r}\right)   &  =\int\frac{\mathbf{r}^{\prime}%
}{r^{\prime}}\frac{\mathbf{r}^{\prime}}{r^{\prime}}\delta\left(  \frac{\sigma
}{2}-r^{\prime}\right)  n\left(  \mathbf{r}-\mathbf{r}^{\prime}\right)
d\mathbf{r}^{\prime}\nonumber
\end{align}
It is sometimes interesting to separate the latter into a sum of its trace
and traceless part as $\mathbf{T}\left(  \mathbf{r}\right)  =\frac{1}%
{3}s\left(  \mathbf{r}\right)  \mathbf{1+U}\left(  \mathbf{r}\right)  $.
Notice that since the density itself is always greater than or equal to zero,
$\eta,s$ and $\mathbf{T}$ are also positive semi-definite. In a liquid, for
which the density is by definition uniform $n\left(  \mathbf{r}\right)
=\overline{n}$, these become%
\begin{align}
\eta\left(  \mathbf{r}\right)   &  =\frac{\pi}{6}\overline{n}\sigma^{3}\\
s\left(  \mathbf{r}\right)   &  =\pi\overline{n}\sigma^{2}\nonumber\\
\mathbf{v}\left(  \mathbf{r}\right)   &  =\mathbf{0}\nonumber\\
\mathbf{T}\left(  \mathbf{r}\right)   &  =\frac{\pi}{3}\overline{n}\sigma^{2}%
\mathbf{1.}\nonumber
\end{align}

Different FMT models are distinguished by the form of $\Phi\left(  n_{\alpha
}\right)  $. In Rosenfeld's original proposal, this was%
\begin{equation}
\beta \Phi_{\text{R}}\left(  n_{\alpha}\right)  =-\frac{1}{\pi\sigma^{2}}s\ln\left(
1-\eta\right)  +\frac{1}{2\pi\sigma}\frac{s^{2}-v^{2}}{\left(  1-\eta\right)
}+\frac{1}{24\pi}\frac{s^{3}-3sv^{2}}{\left(  1-\eta\right)  ^{2}}.
\end{equation}
As long as $\eta<1$, the first term on the right is obviously positive
semi-definite. The second term shares this property as can be seen from the simple
inequality%
\begin{align} \label{inequal}
0  &  \leq\int\delta\left(  \frac{\sigma}{2}-r^{\prime}\right)  n\left(
\mathbf{r-r}^{\prime}\right)  \left(  \widehat{\mathbf{r}}^{\prime}s\left(
\mathbf{r}\right)  -\mathbf{v}\left(  \mathbf{r}\right)  \right)
^{2}d\mathbf{r}\\
&  =\left(  s^{2}\left(  \mathbf{r}\right)  -\mathbf{v}^{2}\left(
\mathbf{r}\right)  \right)  s\left(  \mathbf{r}\right) \nonumber
\end{align}
so $s^{2}\left(  \mathbf{r}\right)  \geq\mathbf{v}^{2}\left(  \mathbf{r}%
\right)  $. On the other hand, the third term is not. As a somewhat realistic
example that illustrates this, consider a hard wall with normal in the
$z$-direction located at $z=0$. Let the density be zero for $z<0$, "outside"
the volume, and a constant $\rho_{0}$ for $z>0$. The weighted densities are
zero for $z<-\sigma/2$, and in the domain $-\sigma/2<z<\sigma/2$ they are (see
Appendix \ref{Calcs} for details)%
\begin{align}
\eta\left(  z\right)   &  =\frac{\pi}{12}\rho_{0}\sigma^{3}\left(  1-\left(
\frac{z}{\sigma}\right)  \right)  \left(  1+2\left(  \frac{z}{\sigma}\right)
\right)  ^{2}\\
s\left(  z\right)   &  =\frac{\pi}{2}\rho_{0}\sigma^{2}\left(  1+2\left(
\frac{z}{\sigma}\right)  \right) \nonumber\\
v\left(  z\right)   &  =\frac{\pi}{4}\rho_{0}\sigma^{2}\left(  1-4\left(
\frac{z}{\sigma}\right)  ^{2}\right) \nonumber
\end{align}
so that%
\begin{align}
s^{2}\left(  z\right)  -v^{2}\left(  z\right)   &  =\frac{1}{8}\left(  \pi
\rho_{0}\sigma^{2}\right)  ^{2}\left(  \frac{3}{2}-\left(  \frac{z}{\sigma
}\right)  \right)  \left(  1+2\left(  \frac{z}{\sigma}\right)  \right)  ^{3}\\
s^{2}\left(  z\right)  -3v^{2}\left(  z\right)   &  =\frac{3}{4}\left(
\pi\rho_{0}\sigma^{2}\right)  ^{2}\left(  1+2\left(  \frac{z}{\sigma}\right)
\right)  ^{2}\left(  \frac{1}{12}+\left(  \frac{z}{\sigma}\right)  -\left(
\frac{z}{\sigma}\right)  ^{2}\right) \nonumber
\end{align}
Clearly, for values of $z$ in the given domain, the first expression is indeed
non-negative. However, the second is not: for example, its value at
$z=-\sigma/4$ is
\begin{equation}
s^{2}-3v^{2}=-\frac{11}{256}\left(  \pi\rho_{0}\sigma^{2}\right)  ^{2}.
\end{equation}
There is no a priori physical reason that this numerator must be
non-negative and it is, in fact, the reason that Rosenfeld's model fails to
stabilize the solid phase: in that application, the numerator can be negative
while the local value of $\eta$ tends to one near the lattice sites in the solid thus making the free energy
unbounded from below. Nevertheless, Rosenfeld's model does give an
impressively good description of a hard-sphere fluid near a wall and has the
important property that it reproduces the Percus-Yevik direct correlation
function for hard-spheres via the exact relation\cite{LutskoAdvChemPhysDFT}%
\begin{equation}
c\left(  r_{12};\overline{n}\right)  =-\lim_{n\left(  \mathbf{r}\right)
\rightarrow\overline{n}}\frac{\delta^{2}\beta F_{ex}\left(  n\right)  }{\delta
n\left(  \mathbf{r}_{1}\right)  \delta n\left(  \mathbf{r}_{2}\right)  }%
\end{equation}
and it follows that the functional also reproduces the Percus-Yevik equation
of state for the uniform fluid. Therefore, much work went into trying to
improve it. The best functionals currently in use are based on Tarazona's
generalization, in which the numerator of the third term is replaced by
\begin{align}
s^{3}-3s\mathbf{v}^{2}  &  \rightarrow\frac{9}{2}\left(  \mathbf{v\cdot T\cdot
v}-sv^{2}+s\operatorname*{Tr}\left(  T^{2}\right)  -\operatorname*{Tr}\left(
T^{3}\right)  \right) \\
&  =s^{3}-3sv^{2}\mathbf{+}\frac{9}{2}\left(  \mathbf{v\cdot U\cdot
v}-\operatorname*{Tr}\left(  U^{3}\right)  \right)  .\nonumber
\end{align}
The resulting functional, $\Phi_{\text{T}}\left(  n_{\alpha}\right)  $, is
indeed able to stablize the solid phase and like Rosenfeld's original
functional, Tarazona's reproduces the Percus-Yevik direct-correlation function
and equation of state for the uniform fluid. It does not, however, give very
good results for liquid-solid coexistence of hard spheres. This is because the
Percus-Yevik approximation is not very good at liquid densities and so the
liquid equation of state is poorly approximated. Roth et al\cite{white_bear} suggested an
empirical modification of the form
\begin{equation} \label{WBI}
\beta \Phi^{\text{WBI}}\left(  n_{\alpha}\right)  =-\frac{1}{\pi\sigma^{2}}%
s\ln\left(  1-\eta\right)  +\frac{1}{2\pi\sigma}\frac{s^{2}-v^{2}}{\left(
1-\eta\right)  }+\frac{3}{16\pi}\frac{\mathbf{v\cdot T\cdot v}-sv^{2}%
+s\operatorname*{Tr}\left(  T^{2}\right)  -\operatorname*{Tr}\left(
T^{3}\right)  }{\left(  1-\eta\right)  ^{2}}\phi_{2}^{\text{WBI}}\left(
\eta\right)  .
\end{equation}
with
\begin{equation}
\phi_{2}^{\text{WBI}}=1-\frac{-2\eta+3\eta^{2}-2\left(  1-\eta\right)  ^{2}%
\ln\left(  1-\eta\right)  }{3\eta^{2}}%
\end{equation}
Whereas Tarazona's functional was based on demanding that FMT reproduce
certain exact limits, this functional, known as the "White Bear" functional, was
mostly an empirical modification aimed at reproducing the more accurate
Carnahan-Starling equation of state for the liquid phase. With this
modification, the theory gives a quantitatively accurate desciption of the
hard-sphere solid and of liquid-solid coexistence. Since the equation of state
for the uniform fluid is no longer that of the Percus-Yevik approximation, it
is necessarily the case that the implied direct correlation function is also
different. In fact, the WBI model gives%
\begin{equation}
c_{WBI}\left(  r;\overline{n}\right)  =\left(  a_{0}^{\left(  WBI\right)
}\left(  \overline{\eta}\right)  +a_{1}^{\left(  WBI\right)  }\left(
\overline{\eta}\right)  r+a_{3}^{\left(  WBI\right)  }\left(  \overline{\eta
}\right)  r^{3}\right)  \Theta\left(  \sigma-r\right)
\end{equation}
with $\overline{\eta}=\frac{\pi}{6}\overline{n}\sigma^{3}$ and%
\begin{align}
a_{0}^{\left(  WBI\right)  }\left(  \eta\right)   &  =\frac{1+4\eta+3\eta
^{2}-2\eta^{3}}{\left(  1-\eta\right)  ^{4}}\\
a_{1}^{\left(  WBI\right)  }\left(  \eta\right)   &  =-\frac{2-\eta+14\eta
^{2}-6\eta^{3}}{\left(  1-\eta\right)  ^{4}}-2\frac{\ln\left(  1-\eta\right)
}{\eta}\nonumber\\
a_{3}^{\left(  WBI\right)  }\left(  \eta\right)   &  =\frac{3-10\eta+15\eta
^{2}-5\eta^{3}}{\left(  1-\eta\right)  ^{4}}+3\frac{\ln\left(  1-\eta\right)
}{\eta}.\nonumber
\end{align}
For reference, we note that the Percus-Yevik direct correlation function has
the same form but with coefficients%
\begin{align}
a_{0}^{\left(  PY\right)  }\left(  \eta\right)   &  =\frac{\left(
1+2\eta\right)  ^{2}}{\left(  1-\eta\right)  ^{4}}\\
a_{1}^{\left(  PY\right)  }\left(  \eta\right)   &  =-\frac{3}{2}\eta
\frac{\left(  2+\eta\right)  ^{2}}{\left(  1-\eta\right)  ^{4}}\nonumber\\
a_{2}^{\left(  PY\right)  }\left(  \eta\right)   &  =\frac{1}{2}\eta
\frac{\left(  1+2\eta\right)  ^{2}}{\left(  1-\eta\right)  ^{4}}\nonumber
\end{align}
The WBI direct correlation function is in fact more accurate than the PY direct correlationh function at high densities\cite{white_bear}.

Finally, a similar functional was developed using a more accurate equation of
state for mixtures, the so-called "White Bear II"\ model\cite{WBII},
\begin{equation} \label{WBII}
\beta \Phi^{\text{WBII}}\left(  n_{\alpha}\right)  =-\frac{1}{\pi
\sigma^{2}}s\ln\left(  1-\eta\right)  +\frac{1}{2\pi\sigma}\frac{s^{2}-v^{2}%
}{\left(  1-\eta\right)  }\phi_{1}^{\text{WBII}}\left(  \eta\right)  +\frac
{3}{16\pi}\frac{\mathbf{v\cdot T\cdot v}-sv^{2}+s\operatorname*{Tr}\left(
T^{2}\right)  -\operatorname*{Tr}\left(  T^{3}\right)  }{\left(
1-\eta\right)  ^{2}}\phi_{2}^{\text{WBII}}\left(  \eta\right)  .
\end{equation}
with
\begin{align}
\phi_{1}^{\text{WBII}}\left(  \eta\right)   &  =1+\frac{2\eta-\eta
^{2}+2\left(  1-\eta\right)  \ln\left(  1-\eta\right)  }{3\eta}\\
\phi_{2}^{\text{WBII}}\left(  \eta\right)   &  =1-\frac{2\eta-3\eta^{2}%
+2\eta^{3}+2\left(  1-\eta\right)  ^{2}\ln\left(  1-\eta\right)  }{3\eta^{2}%
}.\nonumber
\end{align}
It is worth pausing to note that it is easy to show that $\phi_{2}^{\text{WBII}}\left(
\eta\right)  \geq0$ for $0\leq\eta\leq1$. Begin by writing  $\phi_{2}^{\text{WBII}}\left(
\eta\right)  =\frac{\left(  1-\eta\right)  ^{2}}{3\eta^{2}}K\left(
\eta\right)  $ with $K\left(  \eta\right)  =\frac{\eta\left(  -1+3\eta
-\eta^{2}\right)  }{\left(  1-\eta\right)  ^{2}}-\ln\left(  1-\eta\right)  $.
Now $K\left(  0\right)  =0$ and $K^{\prime}\left(  \eta\right)  =\frac{\eta
}{1-\eta}+\frac{2\eta}{\left(  1-\eta\right)  ^{3}}\geq0$ so increases
monotonically from $0$ and is therefore always nonegative and therefore so is
$\phi_{2}^{\text{WBII}}\left(  \eta\right)  $. On the other hand, $\phi
_{2}^{\text{WBII}}\left(  \eta\right)  $ has the well-behaved value $\phi
_{2}^{\text{WBII}}\left(  1\right)  =\frac{2}{3}$ so that it does not affect the postivity of the third term in the $\Phi$ and it introduces no
additional singular behaviour in the high-density limit.

As noted above, it has been reported in the literature that fully
three-dimensional calculations for the hard sphere solid using the tensor
theories must be carefully controled to preserve cubic symmetry or numerical
instabilities are encountered\cite{Oettel}. \red{In our own calculations, we find that the tensorial term does indeed lead to numerical instabilities. In particular, in the solid phase with its highly localized density peaks leading to local packing fractions $\eta(\mathbf{r})$ very close to one, the instabilities are due to the numerator becoming negative leading to an overwhelming large negative contribution to the total free energy. The result is that the total free energy tends to negative infinity and so no physical minimum exists. The question then is whether this is a numerical problem, which can be corrected, or a property of the functional itself. We therefore first ask whether the numerator of the tensorial functional is positive-definite, like the other FMT contributions, or not. If not, this opens the possibility that}  the numerical instabilities are in fact
of the same origin as in Rosenfeld's \red{original} model. (Note however, that in earlier
calculations using Gaussians to represent the solid density, Rosenfeld's model
was unstable but the tensor models were not\cite{tarazona-freezing}.) Returning to the example of the
discontinuous density at a wall, one finds (see Appendix \ref{Calcs} for details) that the tensor density in the domain
$-\sigma/2<z<\sigma/2$ is
\begin{equation}
  T_{ij}\left(  z\right)  = \frac{\pi}{6}\rho_{0}\left(  1-\frac{z}{\sigma}\right)  \left(  2z+\sigma\right)  ^{2}\delta_{ij}
+ \frac{\pi}{2}\rho_{0}\frac{z}{\sigma}\left(
4z^{2}-\sigma^{2}\right) \delta_{iz}\delta_{jz}
\end{equation}
so that the tensor numerator (from the third term in Eqs.(\ref{WBI}) and (\ref{WBII})),  $N=\mathbf{v\cdot T\cdot v}%
-sv^{2}+s\operatorname*{Tr}\left(  T^{2}\right)  -\operatorname*{Tr}\left(
T^{3}\right)  $, becomes
\begin{equation}
N\left(  z\right)  =\frac{\left(  \pi\rho_{0}\sigma^{2}\right)  ^{3}}%
{144}\left(  1+2\left(  \frac{z}{\sigma}\right)  \right)  ^{5}\left(
1-\left(  \frac{z}{\sigma}\right)  \right)  \left(  1+6\left(  \frac{z}%
{\sigma}\right)  -12\left(  \frac{z}{\sigma}\right)  ^{2}+8\left(  \frac
{z}{\sigma}\right)  ^{3}\right)  \allowbreak.
\end{equation}
All of the factors in this expression are positive for $z$ in the given domain
except for the last one which becomes negative for $z<-0.13\sigma$. Another
example is one with zero density outside the wall and a linearlly increasing
density, $\rho_{0}\frac{z}{a}$, for $z>0$ with $a>\sigma/2$. Again, in the
domain $-\sigma/2<z<\sigma/2$ one finds that
\begin{equation}
N\left(  z\right)  =-\frac{1}{442\,368}\pi^{3}\rho_{0}^{3}\frac{\sigma^{9}%
}{a^{3}}\left(  1+2\left(  \frac{z}{\sigma}\right)  \right)  ^{8}\left(
3-176\left(  \frac{z}{\sigma}\right)  +296\left(  \frac{z}{\sigma}\right)
^{2}-192\left(  \frac{z}{\sigma}\right)  ^{3}+48\left(  \frac{z}{\sigma
}\right)  ^{4}\right)  \allowbreak
\end{equation}
which obviously assumes a small but negative value at $z=0$.
In conclusion, these examples serve to show that the third contribution to the tensor functional is not positive definite. \red{Having also tried, without success, to eliminate the numerical instabilities by carefully controlling the numerics (i.e. by modifying the algorithm so as to preserve the inequalities between the weighted densities, by decreasing the lattice spacing and checking for round-off errors in the calculations) we conclude that the mix of empirical and analytic evidence suggests that the functional is indeed unbounded from below and that } the numerical difficulties
encountered when doing free minimization of the solid are due to this fact.

\subsection{A bounded alternative}

A truely bounded (from below)  alternative was in fact proposed heuristically by Rosenfeld et al
prior to the introduction of the tensor densities\cite{RSLT}. The idea was to replace the
numerator in Rosenfeld's theory, $s^{3}-3sv^{2}=s^{3}\left(  1-3v^{2}%
/s^{2}\right)  $ by $s^{3}\left(  1-3v^{2}/s^{2}+3\left(  v^{2}/s^{2}\right)
^{2}-\left(  v^{2}/s^{2}\right)  ^{3}\right)  =s^{3}\left(  1-\left(
v^{2}/s^{2}\right)  \right)  ^{3}$ which, from Eq.(\ref{inequal}),  is now obviously non-negative  and
which agrees with Rosenfeld's theory when $\left(  v^{2}/s^{2}\right)  <<1$ -
e.g. in the liquid limit. The latter property is necessary if one wishes to
still obain the PY direct correlation function for the liquid. There is
another way to motivate this approximation starting with the tensor theory\cite{LutskoAdvChemPhysDFT}.
One could try to replace the traceless tensor $U$ by a combination of the
scalar and vector quantities. Given that it is traceless and must scale
linearlly with the density, any such approximation must have the form
$U^{app}=A\left(  v^{2}/s^{2}\right)  \frac{v_{i}v_{j}-\frac{1}{3}\delta
_{ij}v^{2}}{s}$ where $A(x)$ is an arbitrary scalar function. Evaluating the
tensor numerator with this approximation gives%
\begin{equation}
N=\frac{2}{9}\left(  s^{3}-3sv^{2}\mathbf{+}3A\left(  v^{2}/s^{2}\right)
\frac{v^{4}}{s}-A^{3}\left(  v^{2}/s^{2}\right)  \frac{v^{6}}{s^{3}}\right)  .
\end{equation}
Interestingly, the simplest choice $A\left(  x\right)  =1$ yields the same
positive-definite numerator. In the following, we refer to this as the RSLT functional. 

One can further "upgrade" to the Carnahan-Starling equation of state either by
 making the same approximation for the tensor density in the WBI and
WBII\ functionals or by following the logic of their original derivations but
starting with the positive-definite ansatz for the numerator: both routes
yield the same result which is
\begin{equation}
\beta \Phi\left(  n_{\alpha}\right)  =-\frac{1}{\pi\sigma^{2}}s\ln\left(
1-\eta\right)  +\frac{1}{2\pi\sigma}\frac{s^{2}-v^{2}}{\left(  1-\eta\right)
}\phi_{1}\left(  \eta\right)  +\frac{1}{24\pi}\frac{s^{3}\left(  1-\left(
v^{2}/s^{2}\right)  \right)  ^{3}}{\left(  1-\eta\right)  ^{2}}\phi_{2}\left(
\eta\right)
\end{equation}
with either $\phi_{1}$ and $\phi_{2}$ being the same as in the tensor case
(i.e. the WBI and WBII functions). Furthermore, it turns out that the direct
correlation function for the fluid is then identical to that obtained from the
tensor theories. It therefore seems that the differences between the two are
subtle and probably only important in highly inhomogeneous systems like a solid. The upgraded RSLT model using the WBI functions will be the main alternative discussed in the remainder of this paper and will be referred to as the modified RSLT or mRSLT model. 

\subsection{Calculations: Canonical vs Grand Canonical interpretations}

The goal is to minimize $\Lambda\left[  n;\phi\right]  $ at fixed chemical
potential and external field resulting in a density $n^{\ast
}\left(  \mathbf{r}\right)  $ satisfying%
\begin{equation}
\left.  \frac{\delta F\left[  n\right]  }{\delta n\left(  \mathbf{r}\right)
}\right\vert _{n^{\ast}\left(  \mathbf{r}\right)  }=\mu-\phi\left(
\mathbf{r}\right)  .
\end{equation}
An alternative is to minimize $F\left[  n\right]  $ at constant total particle
number, $N^{\ast}$, by constructing a Lagrangian
\begin{equation}
L_{\lambda}\left[  n\right]  =F\left[  n\right]  +\int\phi\left(
\mathbf{r}\right)  n\left(  \mathbf{r}\right)  d\mathbf{r}-\lambda\left(  \int
n\left(  \mathbf{r}\right)  -N^{\ast}\right)
\end{equation}
and solving the equations
\begin{align}
0  &  =\frac{\delta L_{\lambda}\left[  n\right]  }{\delta n\left(
\mathbf{r}\right)  }=\frac{\delta F\left[  n\right]  }{\delta n\left(
\mathbf{r}\right)  }+\phi\left(  \mathbf{r}\right)  -\lambda\label{EL}\\
0  &  =\frac{\partial L_{\lambda}\left[  n\right]  }{\partial\lambda}=\int
n\left(  \mathbf{r}\right)  -N^{\ast}.\nonumber
\end{align}
Operationally, one could guess a value of $\lambda$ and solve the first
equation for the density, call it $n_{\lambda}\left(  \mathbf{r}\right)  $ and
then vary $\lambda$ until finding a value that satisfies the second equation
(call it $\lambda_{N^{\ast}}$). The resulting density will satisfy
\begin{equation}
\frac{\delta F\left[  n\right]  }{\delta n\left(
  \mathbf{r}\right)  }_{n_{\lambda_{N^{\ast}}}} = \lambda_{N^{\ast}}-\phi\left(  \mathbf{r}\right)
\end{equation}
and comparison to Eq.\ref{EL} shows that this is precisely the local density for a system at chemical potential $\mu = \lambda_{N^{\ast}}$. There are two ways to interpret this result. Rigorously, this means that the local density so obtained is the {\em grand canonical} local density for the system at this chemical potential. The fact that, as a matter of algorithmic convenience, it was obtained by minimizations at constant particle number is irrelevant. However, it is known\cite{Oxtoby_Evans} that for sufficiently large systems, the ensembles become equivalent so that one can, heuristically, take this local density to be that of a canonical system with fixed particle number $N^{\ast}$. Practically, speaking there are in fact methods to evaluate the corrections involved in this identification but in practice they seem to be negligable for systems of 100 or more particles\cite{GWRE, KGV, PhysRevLett.113.238304}.

Note that the fixed-N procedure can be simplified by eliminating the Lagrange multiplier. Multiplying the first of Eqs.\ref{EL} by the density and
integrating, one sees that this particular density also satisfies%
\begin{equation}
0=\int\left(  \frac{\delta F\left[  n_{\lambda_{N^{\ast}}}\right]  }{\delta
n_{\lambda_{N^{\ast}}}\left(  \mathbf{r}\right)  }+\phi\left(  \mathbf{r}%
\right)  \right)  n_{\lambda_{N^{\ast}}}\left(  \mathbf{r}\right)
d\mathbf{r}-\lambda_{N^{\ast}}N^{\ast}%
\end{equation}
so that one can write%
\begin{equation}
0=\frac{\delta F\left[  n_{\lambda_{N^{\ast}}}\right]  }{\delta n_{\lambda
_{N^{\ast}}}\left(  \mathbf{r}\right)  }+\phi\left(  \mathbf{r}\right)
-\frac{1}{N^{\ast}}\int\left(  \frac{\delta F\left[  n_{\lambda_{N^{\ast}}%
}\right]  }{\delta n_{\lambda_{N^{\ast}}}\left(  \mathbf{r}\right)  }%
+\phi\left(  \mathbf{r}\right)  \right)  n_{\lambda_{N^{\ast}}}\left(
\mathbf{r}\right)  d\mathbf{r}%
\end{equation}
This means that $n_{\lambda_{N^{\ast}}}\left(  \mathbf{r}\right)  $ can also
be characterized as the solution to
\begin{equation}
0=\frac{\delta F\left[  n\right]  }{\delta n\left(  \mathbf{r}\right)  }%
+\phi\left(  \mathbf{r}\right)  -\frac{1}{N^{\ast}}\int\left(  \frac{\delta
F\left[  n\right]  }{\delta n\left(  \mathbf{r}\right)  }+\phi\left(
\mathbf{r}\right)  \right)  n\left(  \mathbf{r}\right)  d\mathbf{r}\label{LM}%
\end{equation}
which gives a more practical route that does not involve a Lagrange multiplier
and it is this equation that we actually solve in numerical calculations (when
working at fixed particle number) via the implementation discussed in Appendix
\ref{AuxilliaryField}. So, in this case, one first fixes the lattice parameter
and the number of particles and determines the minimizing density,
$n_{\lambda_{N^{\ast}},a}\left[  \mathbf{r}\right]  $, giving $\Lambda\left(
N^{\ast},a\right)  =\Lambda\left[  n_{\lambda_{N^{\ast}},a}\right]  $. This
then is minimized with respect to either one of the arguments, say $N^{\ast}$,
to get the minimizing value, $N^{\ast\ast}\left(  a\right)  $, and
$\Lambda\left(  a\right)  =\Lambda\left(  N^{\ast\ast}\left(  a\right)
,a\right)  $. At the same time, one gets $\mu\left(  a\right)  =\lambda
_{N^{\ast\ast}\left(  a\right)  }$ which gives the mapping between lattice
parameter and chemical potential. 

\subsection{Validation for the homogeneous hard-sphere crystal}
In order to validate our algorithm and to make contact with previous calculations, we have performed an extensive study of the homogeneous hard-sphere FCC crystalline phase for which  there is no external field and the
crystal extends indefinitely in all directions. Because there are no
long-ranged interactions, it is sufficient to restrict computations to a
single unit cell with periodic boundary conditions. (In fact, for simplicity,
we use a cubic cell rather than the minimal, non-cubic cell for this lattice.)
The minimal cubic unit cell has sides of length $a$ and contains $4$ lattice
points so that the density of lattice points is $\overline{n}_{\text{FCC}%
}=4/a^{3}$. If every lattice site were occupied, this would also be the
physical density of the crystal but in fact, one expects a certain fraction of
unoccupied lattice sites (i.e. vacancies) so that the actual average density
will be $\overline{n}<\overline{n}_{\text{FCC}}$. It is related to the local
density via%
\begin{equation}
\overline{n}=\frac{1}{a^{3}}\int_{\text{cell}}n\left(  \mathbf{r}\right)
d\mathbf{r},%
\end{equation}
where the notation indicates that the integral is restricted to a single unit
cell. The number of vacancies in the unit cell will then be $\overline
{n}_{\text{FCC}}a^{3}-\overline{n}a^{3}=4-\overline{n}a^{3}$ and the vacancy
concentration (ratio of number of vacancies to the number of lattice sites) is
$\overline{c}_{\text{vac}}=1-\overline{n}/\overline{n}_{\text{FCC}}$.
In principle,  one should
allow the lattice parameter to vary during the minimization. Instead, this minimization is divided into two steps: first, a 
minimization at fixed values of the lattice parameter to get $n_{a}^{\ast}\left(\mathbf{r}\right)$ and then a second minimization with of $\Lambda\left[  n_{a}^{\ast
};\phi\right]$ with respect to $a$.

Our implementation of these calculations follows in broad outline previous work: the density field is discretized on a cubic computational lattice and the weighted densities calculated using fast Fourier Transforms. The main novelty of our implementation is that the rather than using the analytic Fourier Transform of the weighting functions as has been done previously, we use a scheme designed to guarantee that the various analytic inequalities between the weighted densities discussed above are rigorously preserved. This may result, for technical reasons discussed in Appendix \ref{Implementation} in somewhat slower convergence (i.e. the need for a finer lattice spacing) but ensures that no artifacts arise due to the FFT technique. A detailed description of the algorithm is given in Appendix \ref{Implementation}.

Because the results of the calculations are rather technical, we have collected most of the results in Appendix \ref{HS} and only briefly discuss the main results here. First, following Ref.\cite{Oettel} we have minimized at fixed-N and fixed lattice parameter with subsequent minimization with respect to the lattice parameter and examined convergence of the algorithm as the lattice parameter is refined. For sufficiently fine lattice spacing (up to 256 points per hard-sphere lattice parameter) we reproduce previous results for the WB models to several significant figures thus validating the implementation. Doing this for many different densities allows us to determine the phase coexistence with the liquid phase. For the WB models, we are again in close numerical agreement with previous calculations.

Concerning the numerical stability of the algorithm, we found, as reported previously, that numerical instabilities arose when we tried to minimize the WB functionals at constant chemical potential. On the other hand, at fixed particle number, we did not find it necessary to maintain strict cubic symmetry during the minimizations and we were able to use much coarser calculational lattices than were possible with previous implementations\cite{OettelPrivate}. We speculate this may be due to the careful treatment of the FFT calculations and the preservation of the analytic inequalities. 

We then extend the calculations to the proposed mRSLT model. In this case, we found that - as expected - we were able to minimize at fixed chemical potential with no instabilities as in the WB models. Also as expected, due to the more heuristic nature of hte model, the results for liquid-solid coexistence were, while reasonable, somewhat poorer. The liquid(solid) packing fractions at coexistence from simulation\cite{Binder} are $0.492(0.545)$,  the WBII model gives $0.495(0.545)$ and the mRSLT yields $0.513(0.546)$. While an error of some $3\%$ is not excessive, there is nevertheless room for improvement.

A final test was to determine the vacancy concentration as a function of the average density. Oettel et al\cite{Oettel} report this as being a rather sensitive test of the models and is one for which the WBII funcitonal is clearly superior to the WBI functional (indeed, WBI does not produce physical results for this quantity because minimization with respect to the lattice spacing is not possible). The mRSLT functional does produce physically reasonable results that compare well with WBII as well as simulation. 

\section{Application to Simple Liquids}
\subsection{cDFT Functional}
To further illustrate the capabilities of a model that is free of divergences
we consider a simple fluid. The development below is generic and can be adapted to any pair potential but for the sake of calculations, we have used  a Lennard-Jones potential,
\begin{equation}
v_{LJ}\left(  r\right)  =4\varepsilon\left(  \left(  \frac{\sigma}{r}\right)
^{12}-\left(  \frac{\sigma}{r}\right)  ^{6}\right)  ,
\end{equation}
which is cutoff at a separation $r_{c}$ and shifted to give the interaction
potential%
\begin{equation}
v\left(  r\right)  =\left\{
\begin{array}
[c]{c}%
v_{LJ}\left(  r\right)  -v_{LJ}\left(  r_{c}\right)  ,\,\,r<r_{c}\\
0,\,\,r\geq r_{c}%
\end{array}
\right.  .
\end{equation}
All results reported here were obtained using a cutoff of $r_{c} = 3 \sigma$.
The potential is separated into a short-ranged repulsive part,
$v_{0}\left(  r\right)  $, and a long-ranged attractive part, $w\left(
r\right)  $, using the WCA prescription\cite{WCA1,HansenMcdonald} according to which $v\left(  r\right)
=v_{0}\left(  r\right)  +w\left(  r\right)  $ with%
\begin{equation}
w\left(  r\right)  =\left\{
\begin{array}
[c]{c}%
v\left(  r_{0}\right)  ,\,\,r<r_{0}\\
v\left(  r\right)  ,\,\,r\geq r_{0}%
\end{array}
\right.
\end{equation}
where $r_{0}=2^{1/6}\sigma$ is the minimum of the potential. An
effective hard-sphere diameter, $d$, is computed using the Barker-Henderson
prescription\cite{BarkerHend,HansenMcdonald},%
\begin{equation}
d_{T}=\int_{0}^{r_{0}}\left(  1-e^{-\beta v_{0}\left(  r\right)  }\right)  dr
\end{equation}
and it should be noted that the effective hard-sphere diameter depends on
temperature but not on the density. The model DFT\ Helmholtz functional is
then constructed as
\begin{equation}
F\left[  n\right]  =F_{HS}\left[  n;d_{T}\right]  +\frac{1}{2}\int%
\int n\left(  \mathbf{r}_{1}\right)  n\left(  \mathbf{r}_{2}\right)
w\left(  r_{12}\right)  d\mathbf{r}_{1}d\mathbf{r}_{2}.
\end{equation}
For a uniform fluid, $n\left(  \mathbf{r}\right)  =\overline{n}$, all of
the Helmholtz free energy assumes akind of generalized van der Waals form,%
\begin{equation}
\frac{1}{V}F\left(  \overline{n}\right)  \equiv f\left(  \overline{n
}\right)  =f_{HS}\left(  \overline{n};d_{T}\right)  +\frac{1}{2}%
a\overline{n}^{2}%
\end{equation}
with%
\begin{equation}
a=\int w\left(  r\right)  d\mathbf{r}.
\end{equation}

To implement this functional, only two changes from the hard-sphere
calculations are necessary. The first is that one must include the extra term
coming from the attractive part of the potential. This is efficiently
evaluated using standard FFT techniques. The second is that care must be taken
with the periodic boundaries.\ Whereas in the hard-sphere case, it sufficed to
take the calculational cell to be the cubic unit cell (and in fact one can
invoke cubic symmetry to reduce this to an octant of the cubic cell, although
we did not do this) the attractive tail will in general have a range greater
than the lattice spacing and so a larger calculational cell must be
used.\ (Again, other, more efficient but more complicated possibilities exploiting cubic symmetry
and not using the minimum image prescription are possible but not pursued here.)

\subsection{Phase diagram}
\begin{figure}[ptb]
\includegraphics[angle=0,scale=0.7]{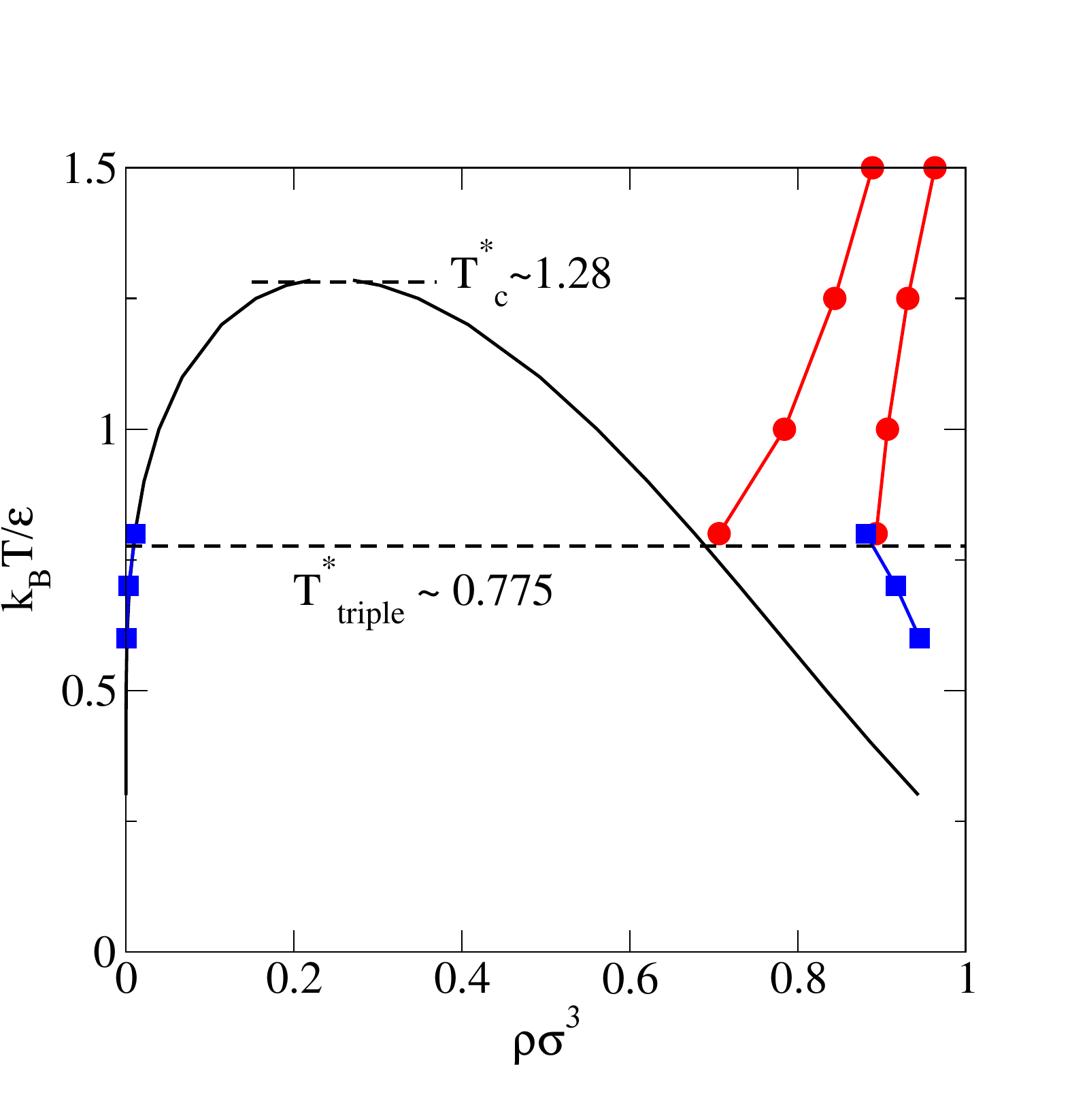} 
\caption{Phase diagram for the Lennard-Jones potential cutoff at $r_{c} = 3\sigma$ as determined from the model DFT functional minimized at constant chemical potential. The dashed lines are for visual purposes only and indicate the critical point and the triple point.}
\label{phasediagram}
\end{figure}

As a first application, the model was used to calculate the vapor-liquid-FCC phase diagram
for a homogeneous system. Since the typical lattice spacing in an FCC Lennard-Jones
solid is on the order of $\sigma$, we used a calculational cell consisting of
$4\times4\times4$ primitive cubic unit cells with a discretization lattice spacing of $\Delta = 0.05\sigma$. The vapor-liquid binodal is easily determined from the model. Indeed, for a uniform system at density $\bar{n}$, one has that
\begin{equation}
  \frac{1}{V} \lim_{n({\bf r}) \rightarrow \bar{n}}\Lambda[n] \equiv \frac{1}{V} \Lambda(\bar{n}) = \left(\bar{n} \ln \left(\bar{n}\sigma^3\right) - \bar{n}\right) +   f^{CS}_{ex}(\bar{n}) + \frac{1}{2}a\bar{n}^2 -\bar{n}\mu
\end{equation}
where the Carnahan-Starling excess free energy per unit volume is
\begin{equation}
  f^{CS}_{ex}(\bar{n})=\bar{n}\frac{\bar{\eta}(4-3\bar{\eta})}{(1-\bar{\eta})^2}
\end{equation}
with $\bar{\eta} = \frac{\pi}{6}\bar{n} d_{T}^{3}$. The Euler-Lagrange equation reduces to
\begin{equation} \label{thermo}
  0 = \frac{\partial \Lambda(\bar{n})}{\partial \bar{n}} \Rightarrow \ln \left(\bar{n}\sigma^3\right)  + \frac{\partial }{\partial \bar{n}}   f^{CS}_{ex}(\bar{n}) + a\bar{n} =  \mu
\end{equation}
which is the usual thermodynamic relation between the chemical potential and the free energy. Coexistence is therefore determined by finding the (two) densities, $\bar{n}_{1}$ and $\bar{n}_{2}$ satisfying Eq.(\ref{thermo}) and, since the fundamental theorem of DFT says the equilibrium state minimizes $\Lambda[n]$, the two states must also satisfy $\Lambda(\bar{n}_{1}) = \Lambda(\bar{n}_{2})$. These two relations - equivalent to the usual thermodynamic criteria of equality of chemical potential and pressure - uniquely determine the coexisting states.

For the solid, the calculation is conceptially the same. First, a chemical potential $\mu$ is chosen. Then, $\Lambda[n]$ is minimized for unit cells of size $k\Delta$ with typical values being $127 \le k \le 135$. Interpolation of the resulting values of $\Lambda$ are used to determine a minimum and, hence, the cell size (i.e. lattice density) and grand-canonical free energy for the solid phase for the given chemical potential. The same interpolation also yields the average density of the solid phase. From Eq.(\ref{thermo}), the density of the liquid (vapor) for the chemical potential is also determined. Finally, the procedure is repeated for different chemical potentials until equality of the grand canonical free energies is found, thus giving the coexisting densities for the solid and liquid (vapor) phases. The resulting phase diagram is shown in Fig.\ref{phasediagram} where the reduced temperature is $T^{\ast} \equiv k_{B}T/\epsilon$. The results are typical of previous calculations and serve as a baseline for the application of this model to crystallization.

\subsection{Heterogeneous Crystallization}
As an example of the capabilities of a well-behaved DFT functional we consider the rapid quench of a small Lennard-Jones droplet attached to a wall. The calculational cell is rectilinear with dimensions $20\sigma \times 20\sigma \times 10\sigma$ with periodic boundaries in the first two (x,y) directions. In the z-direction, the cell is bounded by impenatrable walls. The lower wall at $z=0$ is hard meaning that the particles do not interact with it except that they cannot pass through it: it can be represented as an external potential which is zero for $z>0$ and infinite for $z<0$. The interaction with the upper wall is also taken to be infinite for $z>z_{wall}$ and for $z < z_{wall}$ it is the Steele 4-9 potential\cite{Steele1,Steele2}, which is a continuum potential intended to mimic the average potential generated by the (100) surface of an FCC crystal,
\begin{equation}
  v_{wall}({\bf r})=0.4\epsilon_{wall}\left(\frac{\sigma_{wall}}{z_{wall}-z}\right)^{10}-\left(\frac{\sigma_{wall}}{z_{wall}-z}\right)^{4}-\frac{\sqrt{2}}{3}*\left(\frac{\sigma_{wall}}{z_{wall}-z+0.61\sigma_{wall}/2qrt{2}}\right)^{3}
\end{equation}
where the position of the wall is $z_{wall} = 9\sigma$. (The wall is not at the limit of the cell, $10\sigma$, for technical reasons since the FMT weighted densities are non-zero for one hard-sphere radius outside the calculational domain.) The calculations were performed using $\epsilon_{wall} = 0.15\epsilon$, $\sigma_{wall} = \sigma$ and a discretization of $\Delta = \sigma/7$ which for a typical solid density of $n \sigma^3 = 0.9$ corresponds to approximately $11.5$ lattice points per unit cell. The initial density was taken as  
\begin{equation}
  n({\bf r}) = n_{vap} + (n_{liq}-n_{vap})e^{-\beta v_{wall}({\bf r})}\Theta\left(R - \left({\bf r}_{0}-{\bf r}\right)\right)
\end{equation}
with $n_{vap},n_{liq}$ being the coexistence densities of liquid and vapor at the given temperature, $R$ is the radius of the (spherical) droplet and ${\bf r}_{0} = (10\sigma, 10\sigma, z_{wall})$ is its center. This is therefore a crude model of a spherical drop resting on the upper wall.

The initial temperature was taken to be $k_{B}T=0.8 \epsilon$ and the cDFT functional  minimized at constant particle number. \red{Because of the rather large system sizes needed for the calculations, to ensure that there is no direct self-interaction of the droplets, we have used a rather coarse lattice with a lattice spacing of $\sigma/7$ in this and all subsequent calculations described below.}(There are approximately $297$ particles in the system with almost all concentrated in the droplet.) As discussed above, the resulting configurations can be viewed in two different ways. On the one hand, the droplets are expected to {\em approximate} the density distribution for a canonical system, with the approximation generally becoming more accurate as the number of particles in the system increases. On the other hand, the result of minimizing at constant particle number has the exact interpretation of being the stationary density distribution for the {\em grand canonical system} with a chemical potential determined by the calculation. In the present case, it corresponds to an {\em unstable} stationary point of the grand-canonical functional and, so physically, to the grand-canonical critical droplet.

\begin{figure}[ptb]
\includegraphics[angle=0,scale=0.15]{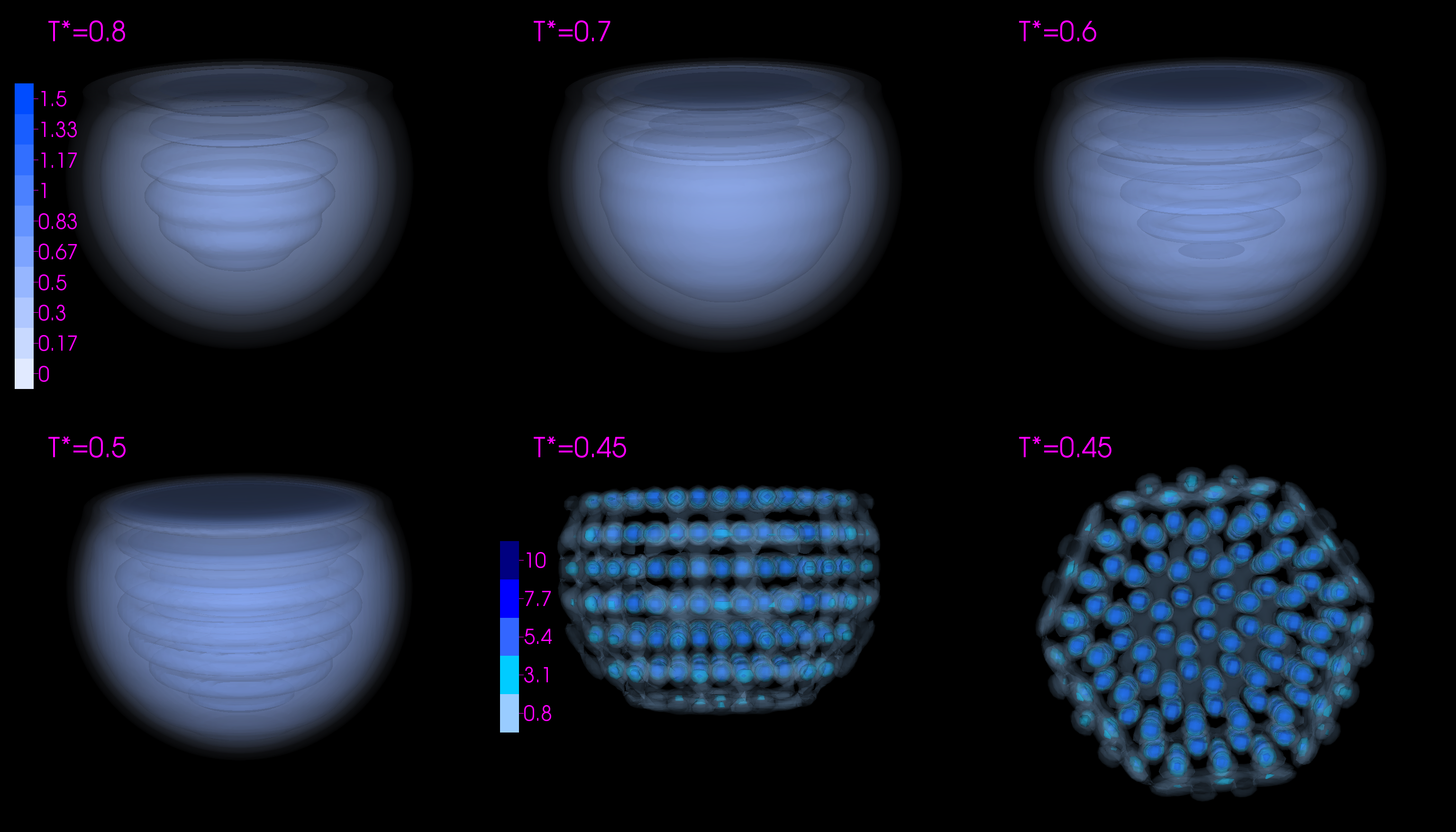} 
\caption{Cooling droplet resulting in crystallization. The images show the density (as a contour plot) with both the intensity and the opacity of the color increasing from very light-blue and completely transparant at zero density to dark blue and 70\% opaque at $n\sigma^{3}=1.5$ for the liquid droplets. For the crystal, the maximum density reaches $216$ and the contours used in the plot are equally spaced from $0.8 \le n\sigma^{3} \le 10$.  The top three images from left to right are at reduced temperatures of $T^{\ast}=0.8, 0.7$ and $0.6$ and $T^{\ast}=0.5$ for the bottom left image. The last two images are different views of the same crystal that spontaneously formed at $T^{\ast}=0.45$.}
\label{droplet}
\end{figure}

Once this configuration is relaxed, the temperature is lowered and the system again relaxed, etc. Since there are no fluctuations in the system (except those induced by numerical noise and the discreteness of the calculational lattice), this corresponds physically to a rapid quench. Figure \ref{droplet} shows the evolution of the density as the temperature is lowered: i.e. according to the two interpretations it is either the evolution of a quenched system in the canonical ensemble or a sequence of critical droplets in the grand canonical ensemble. As the temperature is lowered, one sees the development of more and more structure within the droplet corresponding to layering of the dense fluid near the wall until, at the lowest temperature, the density {\em spontaneously localizes} and the system crystallizes into a hexagonal close-packed (HCP) structure. This crystallization is completely unconstrained and untemplated: it is not induced artificially and the resulting structure is distinct from the cubic calculational lattice. It is also surprising since the minimum energy state for bulk Lennard-Jones is an FCC crystal. However, the difference in energy between HCP and FCC is small and in fact it is known from simulations that small droplets attached to walls will crystallize into HCP structures\cite{Lam}. The crystal can be reheated, with minimization of the energy functional at each temperature, and the free energy difference between the droplet and the crystallite determined until at $k_{B}T=0.55\epsilon$ the crystal spontaneously melts. For the range $0.45 < k_{B}T/\epsilon < 0.55$ both the droplet and the crystallite are metastable thus establishing that there is a free energy barrier separating them: the free energy functional therefore gives a mean-field model of crystallization. At $ k_{B}T/\epsilon=0.50$ the total free energy of the system with crystallite is about $7k_{B}T$ smaller than that with the droplet. Further properties of this model, including a demonstration of the spontaneous formation of the bulk FCC phase, will be discussed in a later publication. 

\subsection{Homogeneous Solidification: Polymorphic Behaviour}

Encouraged by this, we turned to the same problem in the absence of a wall with the aim of studying homogeneous crystals. We began
with a free-floating droplet that was equilibrated at a temperature above the
triple point. We then lowered the temperature and equilibrated resulting in
the series of droplets shown in Figure  \ref{combined_slice}. At a sufficiently low temperature,
we again saw the spontaneous formation of a localized structure. To characterize it, localized peaks obtained in DFT are identified as particle coordinates using particle tracking method that are commonly employed for confocal microscopy analysis\,\cite{Leocmach2013Jan} [See Fig.\ref{DensityVsXYZ}]. From there, we deduced the pair distribution function which exhibit a split of the second peak [See Fig.\,\ref{Structure}.a]. Furthermore, topological cluster classification analysis as developed by Mallins et al.\cite{Malins2013Dec} is carried out in order to identify local structures. The  amorphous cluster is characterized by an increasing emergence of icosahedral order as the temperature is decreased. Icosahedral order has been conjectured as a signature of the glass transition. In this picture, geometrical frustrations are induced because icosahedrons are not able to tessellate 3D space which then lead to the dynamical arrest observed in glass transition\,\citep{Tarjus2005Dec}. While both experimental\cite{Leocmach2012Jul,Pinchaipat2017Jul} and numerical evidences\cite{Wu2016Oct, Hu2015Sep,Turci2017Aug,Turci2017May} have been found for such phenomenon, our results demonstrate that the icosahedral rich structure is not only accompanying or causing the dynamical arrests but it is a genuine minimum in the free energy landscape and should be therefore considered as a thermodynamically stable state. The conclusion that the system has found a free energy minimum corresponding
to a glassy structure is plausible. 
\begin{widetext}
\begin{figure*}[ht]
    \includegraphics[angle=0,scale=0.125]{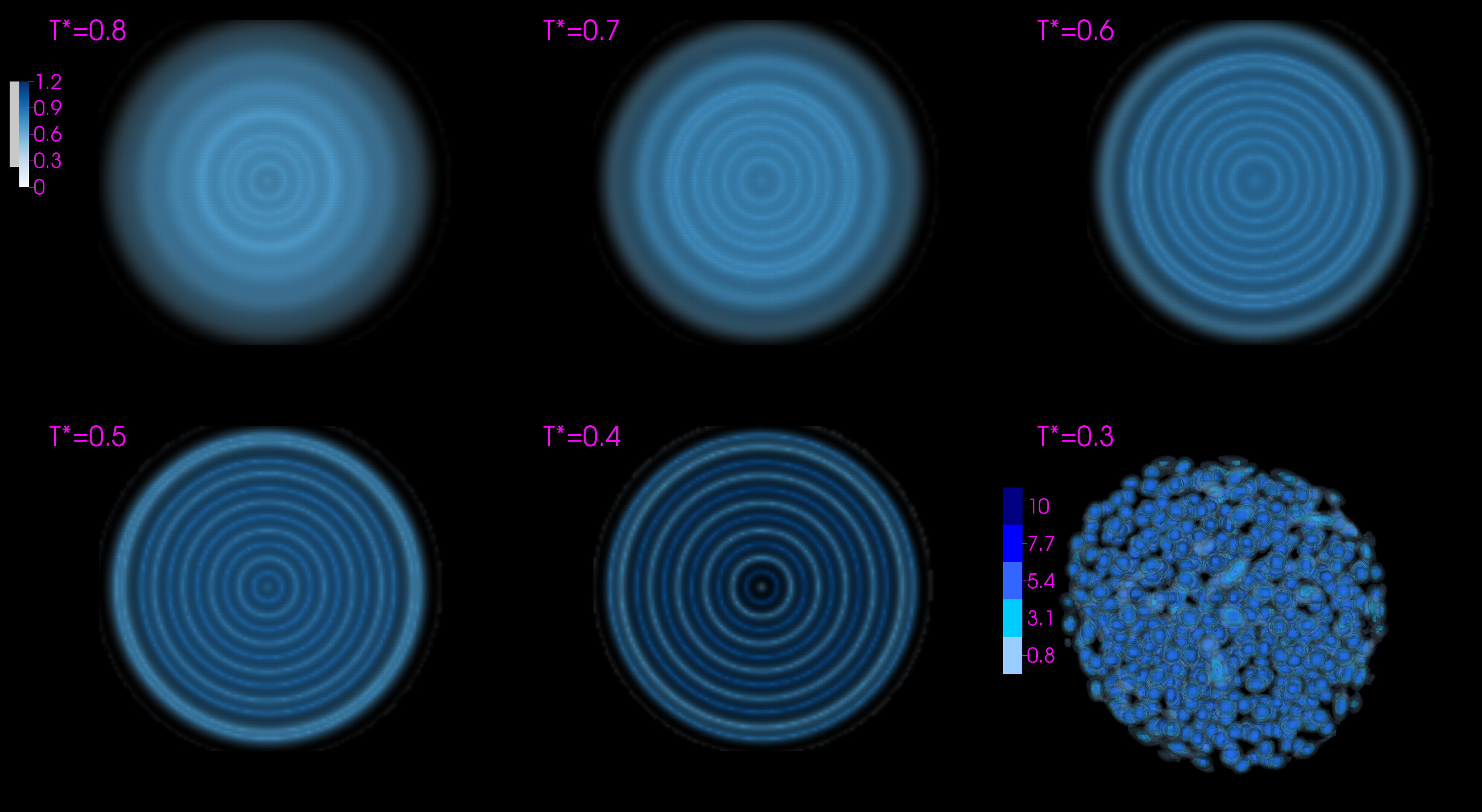}
\caption{The local density of droplets at temperatures $k_{B}T/\epsilon = 0.8, 0.7, 0.6$ from left to right in the first row and $0.5,0.4,0.3$ in the second row: the last figure shows the amorphous cluster that forms spontaneously. For the droplets, the figure displays a planar slice through the center of the droplet with the same color scale, ranging from zero to density $\rho \sigma^{3} = 1.5$. For the amorphous cluster, the representation is a contour plot with the same characteristics as in Fig. \ref{droplet} and with the actual maximum density being $n\sigma^{3}=148$.}
\label{combined_slice}
\end{figure*}
\end{widetext}

\begin{figure}[ht]
    \includegraphics[scale =0.75]{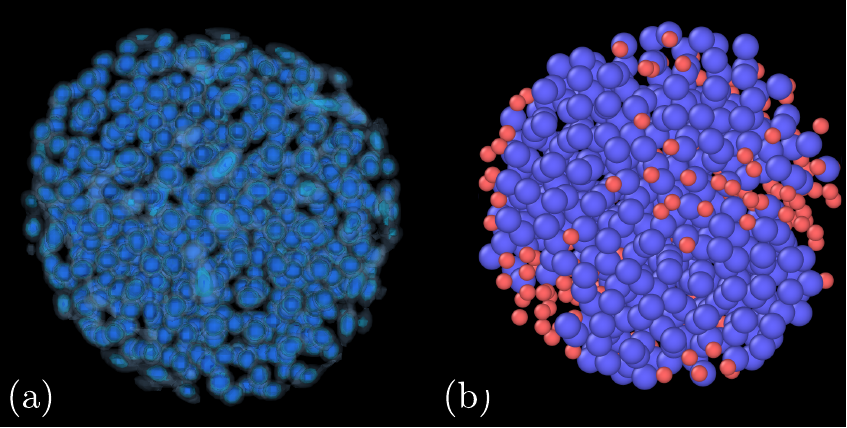}
\caption{(a) Amorphous structure obtained with DFT shown as a contour plot as in Fig.\ref{combined_slice}. (b) Corresponding coordinates identified using particle-tracking. Blue particles are identified as defective icosahedron and others (red) are deliberately shown smaller.}
\label{DensityVsXYZ}
\end{figure}

\begin{figure}[ht]
    \includegraphics[scale =0.75]{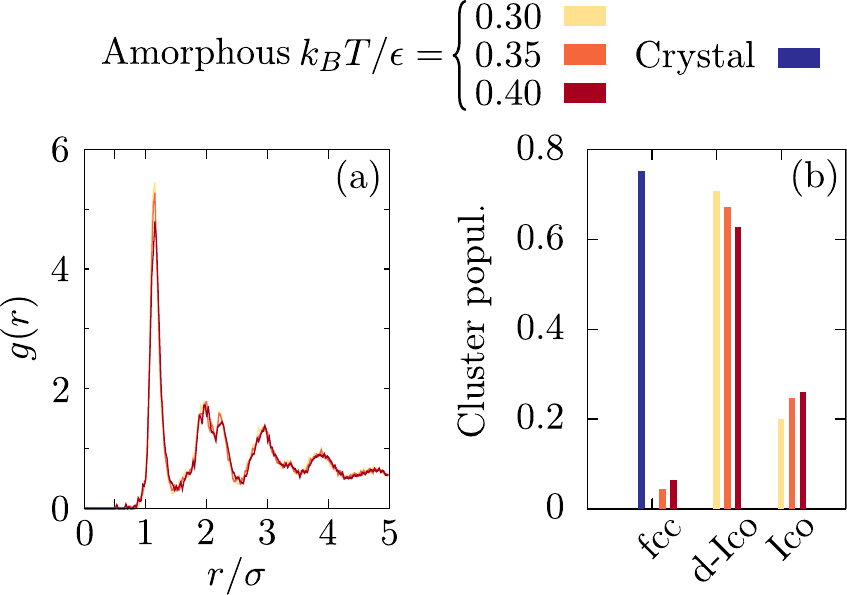}
\caption{(a) Pair distribution function of amorphous structures as obtained at different temperatures. (b) Population of face-centered-cubic, hexagonal compact, defective icosahedron and icosahedron clusters.}
\label{Structure}
\end{figure}

In order to check that this was not a
pathology of the model, we performed the same calculation but at the beginning
of the minimization at the lowest temperature, we introduced a simple planar
field at the center of the droplet which was intended to break the spherical
symmetry. (The specific field was arbitrarily chosen to be $\phi(\mathbf{r})/\epsilon = -0.1 e^{\lvert z \rvert/10}$ where the droplet is centered at $x=y=z=0$. ) This field was removed after a few thousand minimization steps and the
minimization was allowed to continue as before. In this case, we obtained a
nearly perfect FCC crystal with a free energy some $222 k_{B}T$  below that of the
amorphous cluster. The crystal is shown in Fig. \ref{combined_glass_crystal} where its faceting into (111) and (100) planes is evident.  As illustrated in Fig.\ref{Structure}, both hcp and fcc clusters are found within the crystal with far higher populations than for the amorphous cluster.
\begin{widetext}
\begin{figure*}[ht]
    \includegraphics[angle=0,scale=0.15]{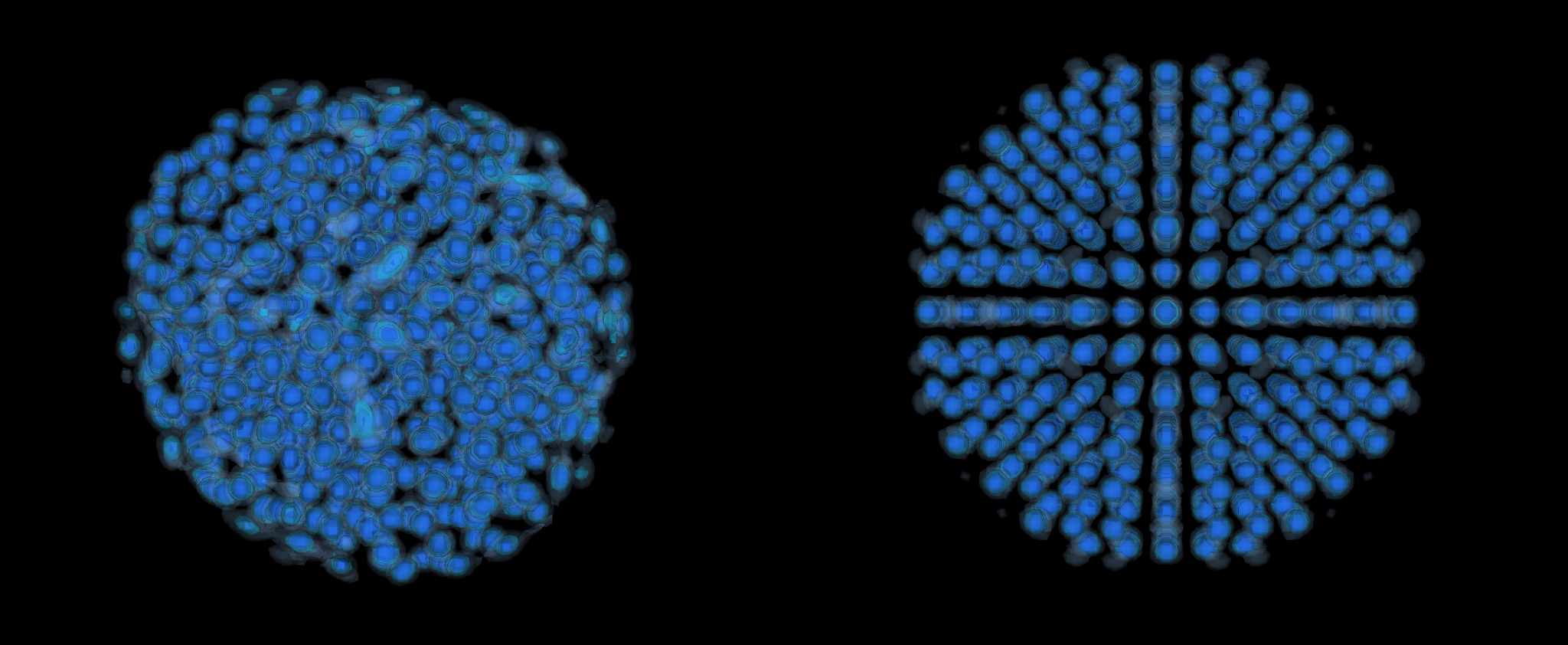}
\caption{A comparison of the  amorphous (left) and crystalline (right) clusters obtained in the homogeneous case, shown as contour plots. Note the faceting of the latter structure. }
\label{combined_glass_crystal}
\end{figure*}
\end{widetext}
From these calculations, we already know that the crystalline and amorphous clusters are both minima of the free energy functional and so there must necessarily be a free energy barrier between them. To investigate their stability, we then increased the temperature and minimized starting in each case with the previous (lower temperature) structure. The resulting free energies are shown in Fig.\ref{FE}. The amorphous structure spontaneously melts at $k_{B}T/\epsilon = 0.45$ indicating that the energy barrier separating it from the liquid state vanishes. The solid is stable up to $k_{B}T/\epsilon = 0.45$ at which point it also melts. 

We note that it is widely known that glasses are not normally seen in simulations of one-component Lennard-Jones systems and so our amorphous clusters might seem suspicious. However, we have verified that using the particle coordinates as input to a (constant-N, canonical) molecular dynamics simulation, the clusters immediately crystallize so that the energy barrier between the amorphous and crystalline states must be very small compared to the thermal energy, thus explaining why they are not normally observed. We also note that our crystalline cluster does not have the geometry of the minimum energy cluster found by Xing et al\cite{Xiang2004Apr} and reported in the Cambridge Cluster Database\cite{CCD}. To investigate this, we have again used our structures as the starting state for the minization of the Hamiltonian in a particle-based algorithm (using facilities in the LAMMPS package\cite{LAMMPS})  and we have performed the same calculation using the database structure (but using our cut and shifted Lennard-Jones potential). We find that the energy per particle of the amorphous cluster converges to $-6.20\epsilon$, our FCC crystal gives $-6.48\epsilon$ and the database structure gives $-6.50\epsilon$ so that while the latter is, technically, a more stable state, the energy difference is very small.


\begin{figure}[ht]
    \includegraphics[angle=0,scale=0.35]{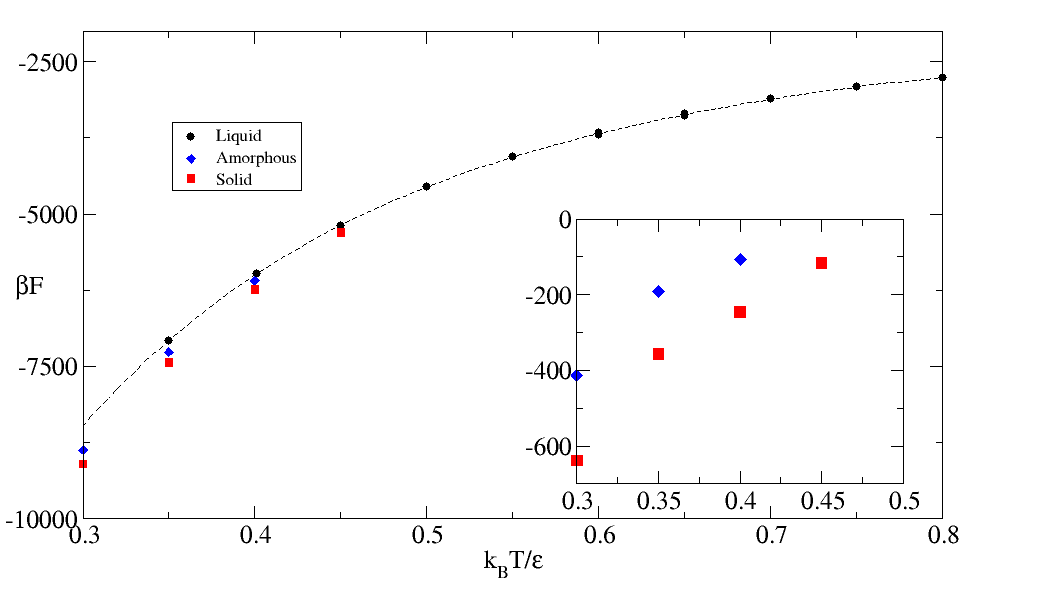}
\caption{The (canoncical, Helmholtz) free energy of the liquid, amorphous and crystalline clusters as functions of the temperature. For the amorphous and crystalline clusters, the curves were obtained by starting with the lowest temperature structures, increasing the temperature and minimizing. The inset shows the free energy of the amorphous and solid clusters relative to the liquid doplet.}
\label{FE}
\end{figure}

Finally, it is important to be clear that our calculations do not, per se, involve nucleation because there are no thermal fluctuations:\ we quench to sufficiently low temperatures that the barrier to crystallization either
vanishes or becomes so small that numerical noise is enough to drive the transition. But such a model can obviously be used together with techniques for exploring free energy surfaces to give a complete picture of
crystallization.  It should be emphasized that DFT by itself is not sufficient to resolve all issues associated with the nucleation of stable solid clusters. This is because nucleation is a fundamentally non-equilibrium process and, has been recently empahsized\cite{Lutsko_JCP_2011_Com,Lutsko_JCP_2012_1}, requires a dynamical description in addition to the free energies provided by DFT. However, it has recently become clear how to combine DFT and dynamics - e.g. fluctuating hydrodynamics - so as to describe first order phase transitions\cite{Lutsko_JCP_2012_1}. The relation between this dynamical description and the more coarse-grained approach of Classical Nucleation Theory has been established as has the development of reduced description based on coarse-grained order parameters\cite{Lutsko_JCP_2012_1,Lutsko2018}. With the addition of free energy functionals capable of describing crystalline phases without constraint, the path is now open to attach theoretical questions that have until now been inaccessible.

\section{Conclusions}

In this paper, the limits of current FMT models for hard-sphere
crystallization have been re-examined. It was demonstrated that the tensor
theories are not positive definite \red{and, based on the numerical evidence, it was suggested that they are indeed not bounded from below thus opening the possibility that } the
hard-sphere crystalline states they predict are in fact metastable states. The
true minima, in this case, consist of some sort of highly localized densities
with local packing fractions approaching unity \red{that give an unbounded negative total free energy becoming lower as the density peaks become more localized}. In
other words, the functionals are unbounded from below. As such, they can only
be viewed as very limited realizations of the ``true'' DFT functional which
should give the (known) stable crystalline state as an absolute minimum of the
functional. The correct behaviour has been illustrated by using an older,
positive-definite, \red{and therefore bounded}, excess free-energy functional that is adapted to give the Carnahan-Starling
equation of state. This functional is certainly not the final word in
hard-sphere functionals as it lacks several advantages of the tenor theories:
in particular, the description of hard-sphere
freezing is not as quantitatively accurate as that of the tensor theories and
(probably related) it lacks consistency with some known exact
results that the tensor theories respect\cite{tarazona-freezing}. Nevertheless, the virtues of being
demonstrably bounded are also very clear: namely, the ability to minimize with no
special regard to maintaining cubic symmetry and even the possibility to
minimize at constant chemical potential with no additional effort.

\red{Ours is not the first work to encounter problems with the tensorial functionals. We note that extensions to non-spherical hard particles have also proven difficult as discussed, e.g. by Wittman et al\cite{Wittman} who observed that some of these problems are solved by using another, older, functional of Tarazona and Rosenfeld\cite{TR} which is also demonstrably bounded from below and which could therefore be an interesting alternative to the RSLT functional used here. In any case, none of this definitively shows that the (newer) tensorial functionals are unphysical. The present work only shows that (a) the numerator of the funcitonal is not positive-definite and (b) that numerical evidence suggests it could be unbounded from below. The second point is supported by the fact that simply replacing the functional by the bounded mRSLT functional with no other change to the calculations gives numerical stability. While our algorithmic implementation is nonstandard and could also be responsible for some of the improved stability, we did find the same instabilities with our algorithm as with older ones.} 

The mRSLT functional was then used as the basis for a mean-field
treatment of a Lennard-Jones system. The full vapor-liquid-solid phase diagram
was computed including full minimization of the density. Then, as a first 
illustation of the novel applications that become possible with a stable
functional, the crystallization of a droplet on an attractive wall was
illustrated. The density of the droplet was calculated (at constant particle
number) for lower and lower temperatures until at sufficiently low
temperature, the droplet crystallzes into an HCP structure. This
localization of the density and crystallization is completely spontaneous and takes place with no
externally-imposed lattice structure. Because these are static calculations,
the crystallization is not the result of nucleation: rather, the temperature
is so low that any energy barrier between the liquid and solid states is so
small that either (a) there is in fact no barrier or (b) there is a barrier
but it is so small that numerical noise in the calculation is enough to allow
the system to overcome it. Nevertheless, despite being performed at constant
particle number, the relevance to nucleation at constant chemical potential
was noted so that the present results do in fact have a bearing on nucleation.

The study of clusters was continued by replacing the wall with periodic boundaries. In this case, an amorphous cluster was produced. We also succeeded in generating a lower energy FCC crystalline cluster by introducing a symmetry-breaking external field in the early stages of minimization and then removing it. These clusters were heated and we demonstrated a region of three-phase coexistence until at higher temperatures the amorphous and crystalline clusters eventually melted. This work continues and we are in the process of studying similar clusters of different sizes. We emphasized that the calculations can be viewed as either approximate (but for such large systems, probably very good) results for a fixed particle number canonical system or as exact results in the grand-canonical ensemble. For the canonical systems, the droplets and clusters are stable states due to the constraint of fixed $N$: if the clusters grow, the density of the vapor outside of them decreases and the system becomes understaturated leading to cluster sublimation while if the clusters shrink, the vapor becomes more highly supersaturated leading to cluster growth\cite{LutskoDuran}. Viewed as exact grand-canonical results, the clusters are metastable critical clusters and as such give a direct indication of the nucleation barrier.

These calculations show that cDFT can be a powerful tool in the study of inhomogeneous classical systems  with the capability of producing unexpected results. It is hoped that the potential of this application will serve to inspire more work on how to improve the FMT theories so that they retain their quantitative accuracy while correcting their weaknesses which seem to be tied to the issue of boundedness of the functional. \red{Promising starting points for such developments could be the older Tarazona-Rosenfeld tensorial functional\cite{TR} discussed also by Wittman et al\cite{Wittman} and the recently proposed scalar functional of Hansen-Goos et al\cite{HansenGoos} built upon a new consistency condition discovered by Santos\cite{Santos}.}  The potential applications to nucleation, pre-melting of surfaces, cluster formation in proteins, wetting, crystallization and perhaps even the glass transition are evident. In particular, they open the door to a complete approach to the description of first order phase transitions in which the free energy surfaces produced can be combined with a stochastic-dynamical framework\cite{Lutsko_JCP_2011_Com,Lutsko_JCP_2012_1} and tools such as the dimer method\cite{Dimer} and the string method\cite{String}, as well as stochastic-processes theory\cite{Hanggi, Talkner, lutsko2014a}, so as to predict nucleation rates, nucleation pathways and multi-step nucleation mechanisms.

\begin{acknowledgments}
It is a pleasure to thank Dominique Maes for useful discussions
and for help in identifying the crystalline structure of the nanocrystals. We
want to thank Martin Oettel for supplying the data shown in Fig.
\ref{fig_vac}, for private discussions concerning his work and for insightful comments on an early draft of this manuscript. We also thank Jim Dufty, Vassily Lubchenko, Serafim Kalliadadasis, John Russo, Paddy Royal and Joshua Robinson for comments on this work. We benefited from the comments of two anonymous referees who suggested related cDFT functionals that could be useful in future work. JFL thanks the European Space Agency (ESA) and the Belgian Federal Science
Policy Office (BELSPO) for their support in the framework of the PRODEX Programme, contract number ESA AO-2004-070. The work of JL was funded by the European Union’s Horizon 2020 research and innovation program  within the AMECRYS project under grant agreement no. 712965.
\end{acknowledgments}

\appendix

\section{Calculations for simple local densities}
\label{Calcs}
\subsection{Generating function formalism}
It is convenient to evaluate the weighted densities using a generating
function formalism\cite{LutskoAdvChemPhysDFT}. For a planar geometry with densities that depend only on
the $z$ coordinate, the generating function
\begin{equation}
\tau\left(  z\right)  =\frac{\pi}{64}\int_{z-\sigma/2}^{z+\sigma/2}\left(
\sigma-2\left(  z^{\prime}-z\right)  \right)  ^{2}\left(  \sigma+2\left(
z^{\prime}-z\right)  \right)  ^{2}n\left(  z^{\prime}\right)  dz^{\prime}%
\end{equation}
can be used to determine all of the weighted densities according to the
relations%
\begin{align}
\eta\left(  z\right)   &  =\frac{4}{\sigma}\frac{\partial}{\partial\sigma}%
\tau\left(  z\right) \\
s\left(  z\right)   &  =2\frac{\partial}{\partial\sigma}\eta\left(  z\right)
\nonumber\\
v_{i}\left(  z\right)   &  =\delta_{iz}\frac{\partial}{\partial z}\eta\left(
z\right) \nonumber\\
T_{ij}\left(  z\right)   &  =\delta_{ij}\frac{2}{\sigma}\eta\left(  z\right)
+\delta_{iz}\delta_{jz}\frac{2}{\sigma}\frac{\partial^{2}}{\partial z^{2}}%
\tau\left(  z\right)  .\nonumber
\end{align}
Defining
\begin{align}
v\left(  z\right)   &  =\frac{\partial}{\partial z}\eta\left(  z\right) \\
W\left(  z\right)   &  =\frac{2}{\sigma}\frac{\partial^{2}}{\partial z^{2}%
}\tau\left(  z\right) \nonumber
\end{align}
the tensor numerator becomes%
\begin{align}
N  &  =\mathbf{v\cdot T\cdot v}-sv^{2}+s\operatorname*{Tr}\left(
T^{2}\right)  -\operatorname*{Tr}\left(  T^{3}\right) \\
&  =v^{2}\left(  \frac{2}{\sigma}\eta+W\right)  -sv^{2}+s\left(  2\left(
\frac{2}{\sigma}\eta\right)  ^{2}+\left(  \frac{2}{\sigma}\eta+W\right)
^{2}\right)  -2\left(  \frac{2}{\sigma}\eta\right)  ^{3}-\left(  \frac
{2}{\sigma}\eta+W\right)  ^{3}.\nonumber
\end{align}
\subsection{Linear local density}
For%
\begin{equation}
n\left(  \mathbf{r}\right)  =n_{0}\left(  1+a\frac{z}{\left(  \sigma/2\right)
}\right)
\end{equation}
with $\left\vert a\right\vert <1$ one finds that
\begin{equation}
\tau\left(  z\right)  =\frac{\pi}{120}\sigma^{4}n_{0}\left(  \sigma
+2az\right)
\end{equation}
giving%
\begin{align}
\eta\left(  z\right)   &  =\frac{1}{30}\pi\sigma^{2}n_{0}\left(
5\sigma+8az\right) \\
s\left(  z\right)   &  =\frac{1}{15}\pi\sigma n_{0}\left(  15\sigma
+16az\right) \nonumber\\
v\left(  z\right)   &  =\frac{4}{15}\pi a\sigma^{2}n_{0}\nonumber
\end{align}
and%
\begin{align}
s^{2}\left(  z\right)  -v^{2}\left(  z\right)   &  =\left(  \pi n_{0}%
\sigma^{2}\right)  ^{2}\left(  1-3a^{2}\right) \\
s^{2}\left(  z\right)  -v^{2}\left(  z\right)   &  =\frac{1}{225}\pi^{2}%
\sigma^{4}n_{0}^{2}\left(  256a^{2}\left(  \frac{z}{\sigma}\right)
^{2}-16a^{2}+480a\left(  \frac{z}{\sigma}\right)  +225\right)  \nonumber \allowbreak
\allowbreak\\
s^{2}\left(  z\right)  -3v^{2}\left(  z\right)   &  =\frac{1}{225}\pi
^{2}\sigma^{4}n_{0}^{2}\left(  256a^{2}\left(  \frac{z}{\sigma}\right)
^{2}-196a^{2}+480a\left(  \frac{z}{\sigma}\right)  +225\right).  \nonumber \allowbreak
\end{align}

\subsection{Step function}
Consider the density $n\left(  \mathbf{r}\right)  =n_{0}\Theta\left(
z\right)  $ corresponding to a system with zero density for $z<0$ and a
constant density $n_{0}$ for $z>a$. In this case, one finds that for
$-\frac{\sigma}{2}<z<\frac{\sigma}{2}$,%
\begin{align}
\tau\left(  z\right)   &  =\frac{\pi}{64}\int_{0}^{z+\sigma/2}\left(
\sigma-2\left(  z^{\prime}-z\right)  \right)  ^{2}\left(  \sigma+2\left(
z^{\prime}-z\right)  \right)  ^{2}n_{0}dz^{\prime}\\
&  =\frac{1}{960}\pi n_{0}\left(  2z+\sigma\right)  ^{3}\left(
6z^{2}-9z\sigma+4\sigma^{2}\right) \nonumber
\end{align}
and%
\begin{align}
\eta\left(  z\right)   &  =\left(  -\frac{1}{12}\pi n_{0}\left(
z-\sigma\right)  \left(  2z+\sigma\right)  ^{2}\right) \\
s\left(  z\right)   &  =\frac{1}{2}\pi\sigma n_{0}\left(  2z+\sigma\right)
\nonumber\\
v\left(  z\right)   &  =\frac{1}{4}\pi n_{0}\left(  \sigma^{2}-4z^{2}\right)
\nonumber\\
T_{ij}\left(  z\right)   &  =\delta_{ij}\left(  -\frac{1}{6}\frac{\pi}{\sigma
}n_{0}\left(  z-\sigma\right)  \left(  2z+\sigma\right)  ^{2}\right)
+\delta_{iz}\delta_{jz}\left(  -\frac{1}{2}\pi\frac{z}{\sigma}n_{0}\left(
\sigma^{2}-4z^{2}\right)  \right) \nonumber
\end{align}
so%
\begin{equation}
N\left(  z\right)  =\frac{1}{144}\frac{\pi^{3}}{\sigma^{3}}\sigma^{5}%
\sigma^{4}n_{0}^{3}\left(  2\left(  \frac{z}{\sigma}\right)  +1\right)
^{5}\left(  -8\left(  \frac{z}{\sigma}\right)  ^{4}+20\left(  \frac{z}{\sigma
}\right)  ^{3}-18\left(  \frac{z}{\sigma}\right)  ^{2}+5\left(  \frac
{z}{\sigma}\right)  +1\right)  \allowbreak
\end{equation}
and%
\begin{align}
s^{2}\left(  z\right)  -v^{2}\left(  z\right)   &  =\frac{1}{8}\left(  \pi
n_{0}\sigma^{2}\right)  ^{2}\left(  \frac{3}{2}-\left(  \frac{z}{\sigma
}\right)  \right)  \left(  1+2\left(  \frac{z}{\sigma}\right)  \right)  ^{3}\\
s^{2}\left(  z\right)  -3v^{2}\left(  z\right)   &  =\frac{3}{4}\left(
\pi n_{0}\sigma^{2}\right)  ^{2}\left(  1+2\left(  \frac{z}{\sigma}\right)
\right)  ^{2}\left(  \left(  \frac{1}{\sqrt{3}}+\frac{1}{2}\right)  -\left(
\frac{z}{\sigma}\right)  \right)  \left(  \left(  \frac{z}{\sigma}\right)
+\left(  \frac{1}{\sqrt{3}}-\frac{1}{2}\right)  \right) \nonumber
\end{align}
In particular,
\begin{equation}
s^{2}\left(  -\frac{\sigma}{4}\right)  -3v^{2}\left(  -\frac{\sigma}%
{4}\right)  =-\frac{11}{256}\left(  \pi n_{0}\sigma^{2}\right)  ^{2}.
\end{equation}

\bigskip
\subsection{Step and linear increase in density}
For the density that is zero for $z<0$ and $n_{0}z/a$ for $z>0$, the generating function for $-\frac{\sigma}{2} < z < \frac{\sigma}{2}$ is
For the linearally increasing density,%
\begin{align}
\tau\left(  z\right)   &  =\frac{\pi}{64}\int_{0}^{z+\sigma/2}\left(
\sigma-2\left(  z^{\prime}-z\right)  \right)  ^{2}\left(  \sigma+2\left(
z^{\prime}-z\right)  \right)  ^{2}n_{0}\frac{z^{\prime}}{a}dz^{\prime}\\
&  =\frac{1}{7680}\frac{\pi}{a}n_{0}\left(  2z+\sigma\right)  ^{4}\left(
4z^{2}-8z\sigma+5\sigma^{2}\right) \nonumber
\end{align}
and%
\begin{align}
\eta\left(  z\right)   &  =\frac{1}{192}\frac{\pi}{a}n_{0}\left(
3\sigma-2z\right)  \left(  2z+\sigma\right)  ^{3}\\
s\left(  z\right)   &  =\frac{1}{8}\frac{\pi}{a}\sigma n_{0}\left(
2z+\sigma\right)  ^{2}\nonumber\\
v_{z}\left(  z\right)   &  =-\frac{1}{12}\frac{\pi}{a}n_{0}\left(
z-\sigma\right)  \left(  2z+\sigma\right)  ^{2}\nonumber\\
W\left(  z\right)   &  =\frac{1}{32}\frac{\pi}{a\sigma}n_{0}\left(
\sigma^{2}-4z^{2}\right)  ^{2}\nonumber
\end{align}
giving%
\begin{equation}
N\left(  z\right)  =-\frac{1}{442\,368}\frac{\pi^{3}}{a^{3}\sigma^{3}}%
\sigma^{8}\sigma^{4}n_{0}^{3}\left(  1+2\left(  \frac{z}{\sigma}\right)
\right)  ^{8}\left(  3-176\left(  \frac{z}{\sigma}\right)  +296\left(
\frac{z}{\sigma}\right)  ^{2}-192\left(  \frac{z}{\sigma}\right)
^{3}+48\left(  \frac{z}{\sigma}\right)  ^{4}\right)  \allowbreak
\end{equation}
and%
\begin{align}
s^{2}\left(  z\right)  -v^{2}\left(  z\right)   &  =\frac{1}{576}\frac{\pi
^{2}}{a^{2}}n_{0}^{2}\left(  5\sigma-2z\right)  \left(  2z+\sigma\right)
^{5}\\
s^{2}\left(  z\right)  -3v^{2}\left(  z\right)   &  =-\frac{1}{192}\frac
{\pi^{2}}{a^{2}}n_{0}^{2}\left(  2z+\sigma\right)  ^{4}\left(
4z^{2}-8z\sigma+\sigma^{2}\right) \nonumber
\end{align}

\section{Calculations at fixed particle number}\label{AuxilliaryField}

An alternative to the introduction of a Lagrange multiplier to fix the
particle number is to introduce an auxilliary field $x\left(  \mathbf{r}%
\right)  $ and to write%
\begin{equation}
n_{N^{\ast}}\left(  \mathbf{r}\right)  =N^{\ast}\frac{x\left(  \mathbf{r}%
\right)  }{\int x\left(  \mathbf{r}^{\prime}\right)  d\mathbf{r}^{\prime}}%
\end{equation}
thus constructing densities with fixed particle numberber. Minimizing
$F\left[  n_{N^{\ast}}\right]  +\int\phi\left(  \mathbf{r}\right)  n_{N^{\ast
}}\left(  \mathbf{r}\right)  d\mathbf{r}$ with respect to $x$ gives%
\begin{equation}
0=\int\frac{\delta F\left[  n_{N^{\ast}}\right]  }{\delta n_{N^{\ast}}\left(
\mathbf{r}^{\prime}\right)  }\frac{\delta n_{N^{\ast}}\left(  \mathbf{r}%
^{\prime}\right)  }{\delta x\left(  \mathbf{r}\right)  }d\mathbf{r}^{\prime
}+\int\phi\left(  \mathbf{r}^{\prime}\right)  \frac{\delta n_{N^{\ast}}\left(
\mathbf{r}^{\prime}\right)  }{\delta x\left(  \mathbf{r}\right)  }d\mathbf{r}%
\end{equation}
and using%
\begin{equation}
\frac{\delta n_{N^{\ast}}\left(  \mathbf{r}^{\prime}\right)  }{\delta x\left(
\mathbf{r}\right)  }=N^{\ast}\frac{1}{\int x\left(  \mathbf{r}^{\prime\prime
}\right)  d\mathbf{r}^{\prime\prime}}\delta\left(  \mathbf{r-r}^{\prime
}\right)  -N^{\ast}\frac{x\left(  \mathbf{r}^{\prime}\right)  }{\int x\left(
\mathbf{r}^{\prime\prime}\right)  d\mathbf{r}^{\prime\prime}}\frac{1}{\int
x\left(  \mathbf{r}^{\prime\prime}\right)  d\mathbf{r}^{\prime\prime}}%
\end{equation}
this becomes%
\begin{equation}
0=\frac{N^{\ast}}{\int x\left(  \mathbf{r}^{\prime\prime}\right)
d\mathbf{r}^{\prime\prime}}\left\{  \frac{\delta F\left[  n_{N^{\ast}}\right]
}{\delta n_{N^{\ast}}\left(  \mathbf{r}\right)  }+\phi\left(  \mathbf{r}%
\right)  -\int\left(  \frac{\delta F\left[  n_{N^{\ast}}\right]  }{\delta
n_{N^{\ast}}\left(  \mathbf{r}^{\prime}\right)  }+\phi\left(  \mathbf{r}%
^{\prime}\right)  \right)  \frac{x\left(  \mathbf{r}^{\prime}\right)  }{\int
x\left(  \mathbf{r}^{\prime\prime}\right)  d\mathbf{r}^{\prime\prime}%
}n_{N^{\ast}}\left(  \mathbf{r}^{\prime}\right)  \right\}
\end{equation}
so that $n_{N^{\ast}}$ satisfies%
\begin{equation}
0=\frac{\delta F\left[  n_{N^{\ast}}\right]  }{\delta n_{N^{\ast}}\left(
\mathbf{r}\right)  }+\phi\left(  \mathbf{r}\right)  -\frac{1}{N^{\ast}}%
\int\left(  \frac{\delta F\left[  n_{N^{\ast}}\right]  }{\delta n_{N^{\ast}%
}\left(  \mathbf{r}^{\prime}\right)  }+\phi\left(  \mathbf{r}^{\prime}\right)
\right)  n_{N^{\ast}}\left(  \mathbf{r}^{\prime}\right)  d\mathbf{r}^{\prime}%
\end{equation}
This is exactly the same as the Eq.(\ref{LM}) derived using a Lagrange multiplier.

\section{Algorithm}
\label{Implementation}
\subsection{Introduction}

This Appendix explains the details of the algorithm used in our
calculations.\ The cDFT functional can be written as
\begin{equation}
\Lambda\left[  n;v\right]  =F^{\left(  id\right)  }\left[  n\right]
+F^{\left(  ex\right)  }\left[  n\right]  +\Lambda^{\left(  V\right)  }\left[
n;v\right]
\end{equation}
with the ideal gas contribution%
\begin{equation}
F^{\left(  id\right)  }\left[  n\right]  =\int\left\{  n\left(  \mathbf{r}%
\right)  \ln\left(  n\left(  \mathbf{r}\right)  \sigma^{3}\right)  -n\left(
\mathbf{r}\right)  \right\}  d\mathbf{r}%
\end{equation}
the excess (FMT) contribution is
\begin{equation}
F^{\left(  ex\right)  }\left[  n\right]  =\int\Phi\left(  \eta\left(
\mathbf{r}\right)  ,s\left(  \mathbf{r}\right)  ,\mathbf{v}\left(
\mathbf{r}\right)  ,\mathbf{T}\left(  \mathbf{r}\right)  \right)  d\mathbf{r}%
\end{equation}
and the contribution due to the external field is%
\begin{equation}
\Lambda^{\left(  V\right)  }\left[  n;v\right]  =\int n\left(  \mathbf{r}%
\right)  \left\{  v\left(  \mathbf{r}\right)  -\mu\right\}  d\mathbf{r.}%
\end{equation}
To evaluate this numerically, we discretize space into a set of points
$\mathbf{R}=i\Delta\widehat{\mathbf{x}}+j\Delta\widehat{\mathbf{y}}%
+k\Delta\widehat{\mathbf{z}}$ where $\Delta$ is the lattice spaceing and
$i,j,k\in%
\mathbb{Z}
$. We assume periodic boundaries so that there are $N_{x}$ unique points in
the $x$-direction, $0\leq i<N_{x}$,etc. The value of the density at lattice
point $\mathbf{R}$ will be denoted $n_{\mathbf{R}}$. The thre contributions
can be evaluated using, e.g., the simplest trap-rule expressions%
\begin{align}
F^{\left(  id\right)  }\left[  n\right]   &  \simeq\sum_{\mathbf{R}}\left\{
n_{\mathbf{R}}\ln\left(  n_{\mathbf{R}}\sigma^{3}\right)  -n_{\mathbf{R}%
}\right\}  \Delta^{3}\\
F^{\left(  ex\right)  }\left[  n\right]   &  =\sum_{\mathbf{R}}\Phi\left(
\eta_{\mathbf{R}},s_{\mathbf{R}},\mathbf{v}_{\mathbf{R}},\mathbf{T}%
_{\mathbf{R}}\right)  \Delta^{3}\nonumber\\
\Lambda^{\left(  V\right)  }\left[  n;v\right]   &  \simeq\sum_{\mathbf{R}%
}n_{\mathbf{R}}\left\{  v_{\mathbf{R}}-\mu\right\}  \Delta^{3}.\nonumber
\end{align}
and the only remaining question is the evaluation of local weighted densities.

\subsection{Weighted Densities}
The weighted densities all have the form of convolutions;%
\begin{equation}
n^{\left(  \alpha\right)  }\left(  \mathbf{r}\right)  =\int w^{\left(
\alpha\right)  }\left(  \mathbf{r-r}^{\prime}\right)  n\left(  \mathbf{r}%
^{\prime}\right)  d\mathbf{r}%
\end{equation}
where the weighting functions, $w^{\left(  \alpha\right)  }\left(
\mathbf{r}\right)  $, are relatively localized: $w^{\left(  \eta\right)  }$ is
only nonzero inside a volume of $\frac{\pi}{6}\sigma^{3}$and the others are
nonzero only on the shell defined by $r=\frac{\sigma}{2}$. So one possibility
would be to directly evaluate them in real space.\ The cost would be on the
order of $N_{x}N_{y}N_{z}\left(  \frac{\sigma}{\Delta}\right)  ^{3}$
operations for $\eta\left(  \mathbf{r}\right)  $ and $N_{x}N_{y}N_{z}\left(
\frac{\sigma}{\Delta}\right)  ^{2}$ for the others. Note that in the latter
case, there is the complication that the domain of integration - the shells of
radius $\sigma/2$ - will, in general, contain no lattice points so this is not
so simple as for the case of the volume integral needed for $\eta$ which can
be approximated by simply summing over the lattice points in the volume.\ An
alternative would be to rewrite the convolution in Fourier space as
\begin{equation}
\widetilde{n}^{\left(  \alpha\right)  }\left(  \mathbf{k}\right)
=\widetilde{w}^{\left(  \alpha\right)  }\left(  \mathbf{k}\right)  n\left(
\mathbf{k}\right)  .
\end{equation}
The analytic Fourier transform of the weight functions is easily calculated
and the that of the density can be obtained by means of Fast Fourier Transform
(FFT) at a cost of on the order of $\left(  N_{x}N_{y}N_{z}\right)  \log
_{2}\left(  N_{x}N_{y}N_{z}\right)  $ operations. The product involves
$\left(  N_{x}N_{y}N_{z}\right)  $ operations and the inverse FFT needed to
get the real-space weighted densities gives a total cost of $2\left(
N_{x}N_{y}N_{z}\right)  \log_{2}\left(  N_{x}N_{y}N_{z}\right)  +\left(
N_{x}N_{y}N_{z}\right)  $ operations. The FFT method is advantageous provided
that
\begin{equation}
\left(  \frac{\sigma}{\Delta}\right)  ^{3}>\log_{2}\left(  N_{x}N_{y}%
N_{z}\right)
\end{equation}
or, if there are the same number of points in each direction, $N<2^{\left(
\frac{\sigma}{\Delta}\right)  ^{3}/3}$. Clearly, the advantage of the FFT
method increases as the lattice is refined but even for as few as 8 points per
hard-sphere diameter the FFT method is advantageous provided the number of
latticepoints in each direction is $N<2^{171}$ .

\subsection{Real space or Fourier space weighting functions?}

Thus, the FFT method must clearly be the method of choice in practical
calculations. However, if implemented as described here, there are potential
problems. If the FT of the weighting functions is determined analytically and
that of the density numerically, then there is no guarantee that the resulting
weighted densities will be non-negative or that the inequalities discussed in
the main text will be respected (e.g. that $s_{\mathbf{R}}^{2}>\mathbf{v}%
_{\mathbf{R}}^{2}$). Since one of our main goals was to investigate the
stability of the models, a more careful implementation of this approach - one
guaranteed to respect these properties - was used. Returning to the real-space
version of the calculation, let us assume that there is some set of points
$\mathbf{r}_{a}$ and weights $u_{a}$ with $0\leq a<M$ that allow us to
approximate an integral over a spherical shell of radius $t$ as
\begin{equation}
\int f\left(  \mathbf{r}\right)  \delta\left(  t-r\right)  d\mathbf{r}%
\simeq\sum_{a=1}^{N}u_{a}f\left(  t\mathbf{r}_{a}\right)  .
\end{equation}
Then, the values of the weighted density $s\left(  \mathbf{r}\right)  $ at the
lattice points could be evaluated as
\begin{equation}
s_{\mathbf{R}}=\sum_{a=1}^{N}u_{a}n\left(  \mathbf{R}+\frac{\sigma}%
{2}\mathbf{r}_{a}\right)  .
\end{equation}
Since we only know the density on the lattice points, the value of the density
on the shell must be evaluated with some sort of interpolation. The only
requirement for the following is that the method of interpolation be linear in
the density: we make the simplest choice of trilinear interpolation. For any
point $\mathbf{r}$ in the domain, a lattice point $\mathbf{S}\left(
\mathbf{r}\right)  =$ $i\left(  \mathbf{r}\right)  \Delta\widehat{\mathbf{x}%
}+j\left(  \mathbf{r}\right)  \Delta\widehat{\mathbf{y}}+k\left(
\mathbf{r}\right)  \Delta\widehat{\mathbf{z}}$can be identified having the
property that
\begin{equation}
i\left(  \mathbf{r}\right)  \Delta\leq r_{x}<\left(  i\left(  \mathbf{r}%
\right)  +1\right)  \Delta,
\end{equation}
etc.\ and the value of the density at $\mathbf{r}$ is estimated by doing
linear interpolation with resepct to the eight corners of this cube. Defining%
\begin{equation}
\Delta_{x}\left(  \mathbf{r}\right)  =\left(  r_{x}-S_{x}\left(
\mathbf{r}\right)  \Delta\right)  /\Delta
\end{equation}
etc., the result is that
\begin{equation}
n\left(  \mathbf{r}\right)  \simeq\sum_{i=0}^{1}\sum_{j=0}^{1}\sum_{k=0}%
^{1}n_{\mathbf{S}\left(  \mathbf{r}\right)  \mathbf{+}\left(  i\Delta
,j\Delta,k\Delta\right)  }B_{ijk}\left(  \mathbf{r}\right)
\end{equation}
with%
\begin{equation}
B_{ijk}\left(  \mathbf{r}\right)  =\left(  \delta_{i0}\Delta_{x}\left(
\mathbf{r}\right)  +\delta_{i1}\left(  1-\Delta_{x}\left(  \mathbf{r}\right)
\right)  \right)  \left(  \delta_{j0}\Delta_{y}\left(  \mathbf{r}\right)
+\delta_{j1}\left(  1-\Delta_{y}\left(  \mathbf{r}\right)  \right)  \right)
\left(  \delta_{k0}\Delta_{z}\left(  \mathbf{r}\right)  +\delta_{k1}\left(
1-\Delta_{z}\left(  \mathbf{r}\right)  \right)  \right)
\end{equation}
Finally, noting that for any lattice vector $\mathbf{R}$, we have that
$\mathbf{S}\left(  \mathbf{R}+\mathbf{r}\right)  =\mathbf{R}+\mathbf{S}\left(
\mathbf{r}\right)  $, it follows that $\Delta_{x}\left(  \mathbf{R}%
+\mathbf{r}\right)  =\Delta_{x}\left(  \mathbf{r}\right)  $ so that
\begin{equation}
n\left(  \mathbf{R}+\mathbf{r}\right)  \simeq\sum_{i=0}^{1}\sum_{j=0}^{1}%
\sum_{k=0}^{1}n_{\mathbf{R+S}\left(  \mathbf{r}\right)  \mathbf{+}\left(
i\Delta,j\Delta,k\Delta\right)  }B_{ijk}\left(  \mathbf{r}\right)
\end{equation}
and%
\begin{equation}
s_{\mathbf{R}}=\sum_{\mathbf{R\prime}}n_{\mathbf{R+R}^{\prime}}\overline
{w}_{\mathbf{R\prime}}^{\left(  s\right)  }%
\end{equation}
with%
\begin{equation}
\overline{w}_{\mathbf{R\prime}}^{\left(  s\right)  }=\sum_{a=1}^{N}\sum
_{i=0}^{1}\sum_{j=0}^{1}\sum_{k=0}^{1}u_{a}B_{ijk}\left(  \frac{\sigma}%
{2}\mathbf{r}_{a}\right)  \delta_{\mathbf{R\prime,S}\left(  \frac{\sigma}%
{2}\mathbf{r}_{a}\right)  \mathbf{+}\left(  i\Delta,j\Delta,k\Delta\right)  }.
\end{equation}
Analogous expressions for the other shell-averaged densities are easily found.
For the local packing fraction, we obtain something similar by integrating
this shell expression with respect to the radial variable using a
one-dimensional Gauss-Legendre scheme. These real-space, discretized
expressions for the weighted densities satisfy all of the physical
inequalities and the discrete convolutions can be efficiently (and, up to
numerical noise, exactly) evaluated using discrete FFT.\ 

\subsection{Spherical Integrals}

One question not addressed so far is what points are used to do the spherical
integrals. If one naively applies one-dimensional integration schemes to
spherical coordinates, its easy to see that spatial asymmetries are created
that could bias the calculations. The standard method\cite{Quadrature}, used for example
in quantum-chemical calculations, are to use points and weights some subset of
spherical harmonics exactly. The two most common schemes are Lebedev
quadrature and Spherical t-schemes\cite{Quadrature}. The difference between the two is that the
former strictly enforce octahedral and inversion symmetry whereas the latter
are constructed by demanding that the weights of all points are equal. The
efficency of the two schemes is similar\cite{Quadrature}. Both have been tested and the
differences in the present calculations are negligable. When using spherical
t-schemes, we have enforced cubic and reflection symmetry by expanding the
list of points first so that besides each point $(x,y,z)$ we also have all
permutations of the same values of $x,y,z$ (thus going from one to 6 points).
We then take this expanded list of points and include all reflections:
$(x,y,z), (-x,y,z) ... (-x,-y,z) ... (-x,-y,-z)$. In total, this expands the
list of $N$ points to $N\times6 \times8 = 48N$. In our calculations, we have
used point sets from the website Ref.\cite{Womersley,Rob_Womersley}. The results seem to
be robust with respect to the number of points used: we have typically used
sets containing 21,118 points. The volumetric spherical integrals are
performed by supplementing the integration on the spherical shell with
Gauss-Legendre integration in the radial coordinate using up to 512 points. Numerical calculations were implemented using the GSL scientific library\cite{GSL} and the Armadillo linear algebra library\cite{Armadillo} and  Fourier transforms perfomed using the FFTW library\cite{FFTW}.

\subsection{Minimization}
After discretization, the ftDFT functional becomes a function of the densities at each lattice point. The gradients of this function are easily calculated ``analytically'' (i.e. analytic expressions that are calculated at the same time as the functional itself). Indeed, the discretized $\Lambda$ functional has the form
\begin{equation}
\Lambda({\bf n}) = \sum_{i}(n_{i}\ln (n_{i}\sigma^3) - n_{i}) \Delta^3 + \sum_{i} \Phi({\bf n}) \Delta^3 + \frac{1}{2}a \sum_{i,j} n_{i}n_{j} w_{ij} \Delta^6 + \sum_{i}n_{i}(v_{i}^{ext}-\mu)\Delta^3
\end{equation}
where the sums are over all lattice points, $w_{ik}$ is the attractive part of the potential evaluated at the lattice sites, $v^{ext}_{i}$ is the  external potential evaluated on the lattice sites and ${\bf n}$ is the collection of values of the density on the lattice points. Minimization requires that 
\begin{equation}
0 = \frac{\partial \Lambda({\bf n})}{\partial n_{k}} = \ln (n_{k}\sigma^3) \Delta^3 + \sum_{i} \frac{\partial}{\partial n_{k}}\Phi({\bf n}) \Delta^3 + a \sum_{i} n_{i} w_{ik} \Delta^6 +v_{k}^{ext}-\mu \Delta^3
\end{equation}
or, upon rearrangement,
\begin{equation}
n_{k}\sigma^3   = \exp\left(- \sum_{i} \frac{\partial}{\partial n_{k}}\Phi({\bf n})  - a \sum_{i} n_{i} w_{ik} \Delta^3 -v_{k}^{ext}+\ mu \right)
\end{equation}
These forces were used to implement either conjugate gradient (CG) minimization\cite{NR} or the Fast Inertial Relaxation Engine (FIRE) of Bitzek et al\cite{Bitzek}. For the liquid, both methods acheived convergence quickly and with similar effort. For the solid phase, the CG would sometimes fail to converge or converge very slowly while FIRE always converged and generally with much less effort. 

Minimization was terminated using the criterion
\begin{eqnarray}
  \epsilon & > & \max_{k} \left| \frac{\exp\left(- \sum_{i} \frac{\partial}{\partial n_{k}}\Phi({\bf n})  - a \sum_{i} n_{i} w_{ik} \Delta^3 -v_{k}^{ext} +\mu \right)}{n_{k}\sigma^3}-1 \right| \\
  & = &  \max_{k} \left| \exp\left(-\ln(n_{k}\sigma^3) - \sum_{i} \frac{\partial}{\partial n_{k}}\Phi({\bf n})  - a \sum_{i} n_{i} w_{ik} \Delta^3 -v_{k}^{ext} +\mu \right) -1 \right| \nonumber \\
  & \approx &  \max_{k} \left| \ln(n_{k}\sigma^3) + \sum_{i} \frac{\partial}{\partial n_{k}}\Phi({\bf n})  + a \sum_{i} n_{i} w_{ik} \Delta^3 + v_{k}^{ext}-\mu \right| \nonumber
\end{eqnarray}
where the last approximation is valid when the forces are small (the case of interest) and avoids the expense of exponentiating over all of the lattice points. We found that convergence was generally achieved for $\epsilon = 10^{-2}$  and, to be sure, the criterion $\epsilon = 10^{-4}$ was used throughout the calculations.

\section{Results for hard-spheres}
\label{HS}

\subsection{Convergence and comparison to previous results}

We have implemented these algorithms by discretizing the density on a cubic grid
(details are given in Appendix \ref{Implementation}). Aside from the physical
parameters, the only numerical parameter is the grid spacing, $\Delta$. The convergence of the calculations as the lattice spacing is decreased (at fixed particle number and using
the WBII\ model) is illustrated by the results in Table \ref{table1} which includes an
extrapolation to the continuum limit. Even the coarsest grid, with only $8$
points in each Cartesisan direction, corresponding to about 5 points per hard
sphere diameter, gives free energies accurate to about $6\%$. At $256^{3}$
grid points, the results are within $0.2\%$ of the continuum limit. Compared
to Ref.\cite{Oettel}, which are presumeably converged to several significant
figures, the present calculations converge more slowly. Most likely this is due to 
specific design goals in our algorithm described in Appendix
\ref{Implementation} which lead to the use of linear interpolation and simple
trap-rule integration.

Table \ref{table2} shows results from calculations using the WBII
functional at constant particle number at different densities. In each case, the average density,
$\bar{n}$, and the vacancy concentration, $c_{vac}$, was fixed thus
determining the total number of particles in a unit cell as $N=4
\times(1-c_{vac})$ and thus the lattice parameter via $4(1-c_{vac})/a^{3} =
\bar{n}$. Varying the vacancy concentration while holding the average
density constant is therefore equivalent to varying the lattice parameter. Also shown are 
 results reported in Ref.\cite{Oettel}, where the authors state
that they used $64^{3}$ to $256^{3}$ grid points per unit cell (increasing as
the density increases) and presumeably, the numbers given are converged to a
several significant digits. The present calculations using $256^{3}$ grid points are
consistent with these values with the typical difference being on the order of
$0.1\%$.

\begin{table}
\begin{minipage}{350pt}
\caption{Hard sphere free energies for density $\bar{n} = 1.04086$ (packing fraction $\bar{\eta} = 0.545$) and vacancy concentration $c_{vac} = 1 \times 10^{-4}$ as a function of lattice spacing. The first column indicates the source of the data (WBII are the present calculations), the second the number of lattice points in each Cartesian direction of the unit cell and the third the Helmholtz free energy per particle reduced by the temperature.}
\label{table1}
\addtolength\tabcolsep{2pt}
\begin{tabular}{lll}
\hline \hline
Theory & $a/\Delta$  & $\beta F/N$ \\
\hline
WBII   & $8$    & $5.273$   \\
& $16$    & $5.311$   \\
& $32$    & $5.075$   \\
& $64$    & $5.003$   \\
& $128$   & $4.983$   \\
& $256$   & $4.979$   \\
& $\infty$& $4.968(7)$ \\
WBII (Ref \cite{Oettel}) & $64$  & $4.977$   \\
Simulation(Ref \cite{Oettel})     &         & $4.959$   \\
\hline \hline
\end{tabular}
\end{minipage}

\end{table}

\begin{table}
[ptb]\begin{minipage}{350pt}
\caption{Hard sphere free energies: comparison of present calculations (marked WBII) to Oettel et al. Following Ref. \cite{Oettel}, these calculations were performed at fixed particle number with a constant vacancy concentration of $c_{vac} = 1 \times 10^{-4}$. The columns are the lattice density, the lattice packing fraction, the number of points in each Cartesian direction of the unit cell and the Helmoltz free energy per particle reduced by the temperature. All calculations were performed using the WBII model.}
\label{table2}
\addtolength\tabcolsep{2pt}
\begin{tabular}{llllll}
\hline \hline
$\bar{n}_{latt}\sigma^3$ & $\eta_{latt}$ & Source & $a/\Delta$ & $\beta F/N$\\
\hline
$1.00$    &  $0.5236$ & WBII    & $64$  & $4.558$\\
&& WBII    & $256$ & $4.541$\\
&& Ref. \cite{Oettel}  & $64$  & $4.539$\\
$1.04086$ & $0.5450$  & WBII    & $64$  & $5.003$\\
&& WBII    & $256$ & $4.979$\\
&& Ref. \cite{Oettel}  & $64$  & $4.977$\\
$1.049$   & $0.5493$  & WBII    & $64$  & $5.094$\\
&& WBII    & $256$ & $5.069$\\
&& Ref. \cite{Oettel}  & $64$  & $5.067$\\
$1.08$    & $0.5655$  & WBII    & $64$  & $5.457$\\
&& WBII    & $256$ & $5.424$\\
&& Ref. \cite{Oettel}  & $64$  & $5.422$\\
$1.09975$ & $ 0.5758$ & WBII    & $64$  & $ 5.701$\\
&& WBII    & $256$ & $5.661$\\
&& Ref. \cite{Oettel}  & $64$  & $5.658$\\
$1.11$    & $0.5812 $ & WBII    & $64$  & $5.831$\\
&& WBII    & $256$ & $5.788$\\
&& Ref. \cite{Oettel}  & $64$  & $5.785$\\
$1.14$    & $0.5969$  & WBII    & $64$  & $6.233$\\
&& WBII    & $256$ & $6.176$\\
&& Ref. \cite{Oettel}  & $256$ & $6.172$\\
$1.15$    & $0.6021$  & WBII    & $64$  & $6.375$\\
&& WBII    & $256$ & $6.313$\\
&& Ref. \cite{Oettel}  & $256$ & $6.308$\\
\hline \hline
\end{tabular}
\end{minipage}

\end{table}

\subsection{Coexistence}

Table \ref{table3} shows the coexistence conditions predicted by the various theories.
While the values are somewhat dependent on the grid spacing, they
are largely insensitive to the vacancy concentration. The values of the
postive-definite (mRSLT) theory are also given. Even though the latter
incorporates the Carnahan-Starling equation of state, and is therefore
equivalent to the WBI theory for the liquid phase, it is less accurate in
predicting coexistence due to higher overall predictions for the solid pressure. It therefore lacks the qualitative accuracy of the tensor theories.

\begin{table}
[ptb]\begin{minipage}{350pt}
\caption{Liquid-FCC solid coexistence (WBI, WBII and mRSLT are the present calculations) for hard spheres. }
\label{table3}
\addtolength\tabcolsep{2pt}
\begin{tabular}{lllllll}
\hline \hline
Source & $a/dx$ & $c_{vac}$ & $\eta_{liq}$ & $\eta_{sol}$ & $\beta
P\sigma^{3}$$\beta \mu$\\
\hline
WBII        & $64$     &  $0.005$  & $0.499$ & $0.545$ & $12.26$ & $16.80$\\
WBII        & $256$    &  $0.005$  & $0.496$ & $0.544$ & $12.00$ & $16.52$\\
WBII        & $64$     &  $0.0001$ & $0.497$ & $0.545$ & $12.14$ & $16.66$\\
WBII        & $256$    &  $0.0001$ & $0.495$ & $0.545$ & $11.89$ & $16.40$\\
WBII (Ref. \cite{Oettel}) & $64$     &  $0.0001$ & $0.495$ & $0.544$ & $11.89$ & $16.40$\\
WBI         & $256$    &  $0.005$  & $0.490$ & $0.536$ & $11.41$ & $15.78$\\
WBI         & $256$    &  $0.0001$ & $0.489$ & $0.537$ & $11.30$ & $15.89$\\
WBI (Ref. \cite{Oettel})  & $64$     &  $0.0001$ & $0.489$ & $0.535$ & $11.28$ & $15.75$\\
mRSLT       & $256$    &  $0.0005$ & $0.512$ & $0.548$ & $13.75$ & $18.51$\\
mRSLT       & $256$    &  $0.0001$ & $0.513$ & $0.546$ & $13.92$ & $18.34$\\
Simulation\cite{Binder}          &  NA      &   NA      & $0.492$ & $0.545$ & $11.58$ & \\
\hline \hline
\end{tabular}
\end{minipage}
\end{table}

\subsection{Vacancy concentration}

The results for freezing are relatively insensitive to the vacancy
concentration. Nevertheless, DFT formally requires that the $\Lambda
$-functional be fully minized with respect to the density. In the present
context, this means that it must be minimized with respect to the vacancy
concentration. As discussed in detail in Ref. \cite{Oettel}, such a
minimization is not even possible with functionals such as WBI and the fact
that WBII can be minimized and gives physically reasonable results is one of
its strongest features. In the case of the RSLT functional, there is no doubt
that it can be minimized with respect to the average density (at fixed lattice
constant): this follows simply because as the local density increases, it must
at some point cause the local packing fraction to increase beyond unity and
this will necessarily lead to a (positive) divergence in the functional.
Figure \ref{fig_vac} shows the results of such a minimization, together with
the WBII results and data from simulations. The mRSLT theory gives
physically reasonable results at all densities.

\begin{figure}[ptb]
\includegraphics[angle=0,scale=0.5]{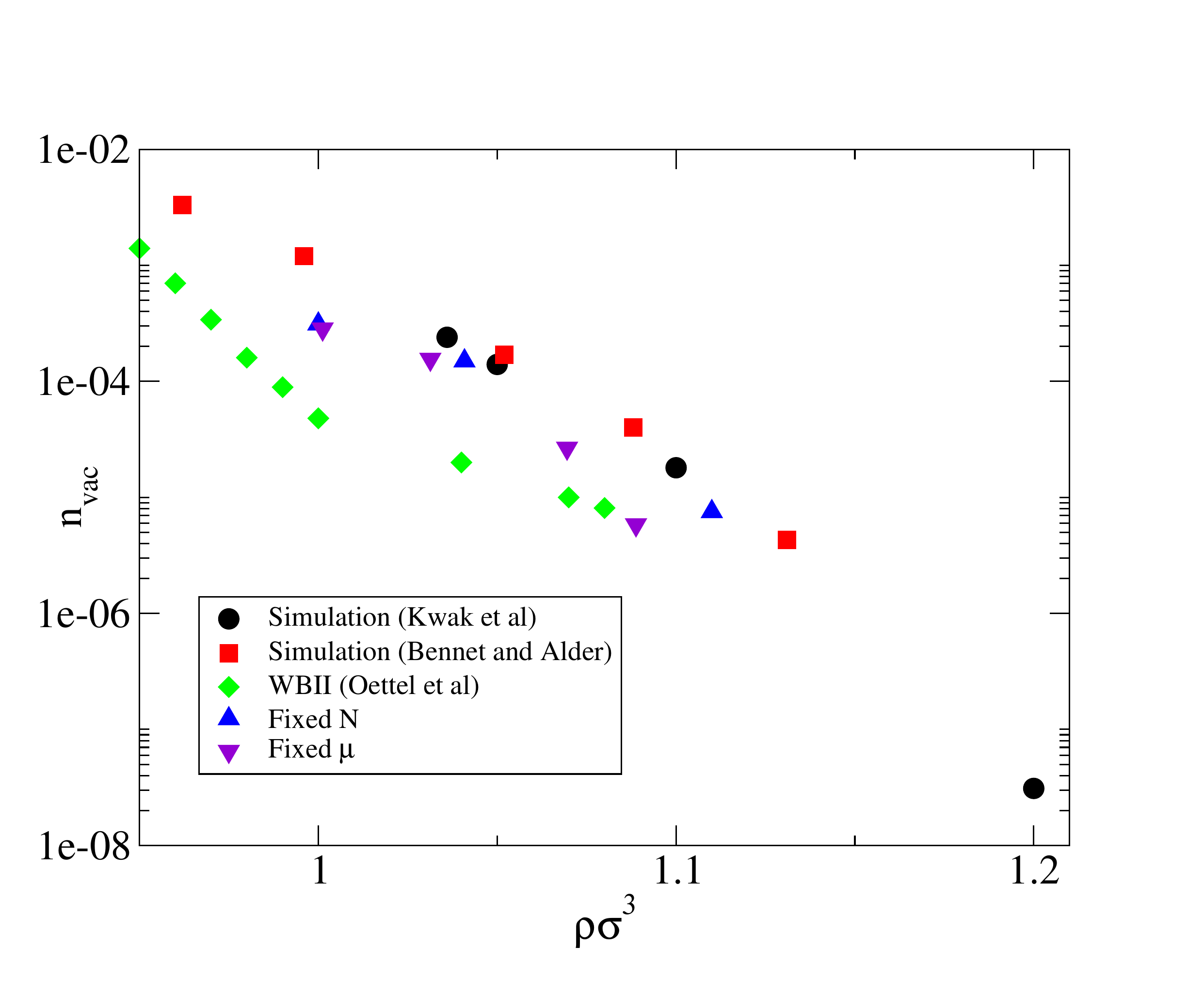} 
\caption{Vacancy concentration as a function of average density determined using the mRSLT model both by minimizing the free energy at constant particle number and at constant chemical potential with subsequent minimization with respect to the lattice spacing in both cases. Simulations results due to Kwak et al\cite{Kwak} and Bennett and Alder\cite{Bennett} are shown as are the WBII results of Oettel et al\cite{Oettel}. }
\label{fig_vac}
\end{figure}

Even more impressive in this context is the fact that there are no technical
difficulties in minimizing at constant chemical potential rather than constant
average density (or, equivalently, constant vacancy concentration). Even for
WBII, Oettel et al state that such a minimization at constant chemical
potential is not technically feasible and we have confirmed with the present
code that attempts to do so typically lead to (negative) divergence of the
energy functional. This is not the case with the mRSLT model and
the Fig. \ref{fig_vac} shows that the resulting vacancy concentrations are consistent with those obtained from the constant density minimizations.

%

\end{document}